\documentclass[fleqn,usenatbib]{mnras}

\usepackage{newtxtext,newtxmath}

\usepackage[T1]{fontenc}

\usepackage{graphicx}	
\usepackage{amsmath}	
\usepackage{lineno}

\usepackage{url}
\usepackage{booktabs}
\usepackage{multirow}
\usepackage{siunitx}
\newcommand{\hb}{{\sc{H}}$\beta$\/}
\newcommand{\feii}{{\ion{Fe}{II}}\/}
\newcommand{\feiiUV}{{\ion{Fe}{II} UV}\/}
\newcommand{\civ}{{\ion{C}{IV}}\/}
\newcommand{\oiii}{{[\ion{O}{iii}]}\/}
\newcommand{\kms}{km s$^{-1}$}

\newcommand{\mgiionly}{\ion{Mg}{II}}
\newcommand{\mgii}{\ion{Mg}{II} $\lambda$2800  \AA\/}
\newcommand{\rfe}{$R_{\rm FeII}$}
\newcommand{\rfeUV}{$R_{\rm FeII}$ UV}

\newcommand{\FWHMmg}{FWHM$_{\ion{Mg}{II}}$}

\newcommand{\TB}[1]{\textbf{#1}}

\title[BZQ Blazars in the E1 UV Main Sequence]{Unveiling the Eigenvector 1 Sequence in BZQ Blazars. I. \TB{Mg II and Ultraviolet Fe II Emission in the UV Main Sequence}}
\author[Buendia-Rios et al.]{
T. M. Buendia-Rios $^{1,2}$
\thanks{SECIHTI Postdoctoral Fellow}
, V. Chavushyan $^{1}$,
V. M. Patiño-Álvarez $^{1,3}$,
J. U. Guerrero-González $^{1}$,
\newauthor M. Martínez-Burelo $^{1}$
and C. A. Negrete$^{2}$
\thanks{SECIHTI Research Fellow}
\thanks{Corresponding author e-mail: alenka@astro.unam.mx}
\\
$^{1}$Instituto Nacional de Astrofísica, Óptica y Electrónica, Luis Enrique Erro $\#$1, Tonantzintla, Puebla 72840, México \\
$^{2}$Universidad Nacional Autónoma de México, Instituto de Astronomía, A.P. 70-264, Ciudad de México, CDMX 04510, México \\
$^{3}$Max-Planck-Institut für Radioastronomie, Auf dem Hügel 69, D-53121 Bonn, Germany \\
}

\date{Accepted XXX. Received YYY; in original form ZZZ}

\pubyear{\the\year{}}

\begin{document}

\label{firstpage}
\pagerange{\pageref{firstpage}--\pageref{lastpage}}
\maketitle

\begin{abstract}
We investigate the location of flat-spectrum radio quasars confirmed as blazars (BZQ-C), Fanaroff–Riley type II radio-loud (RL) quasars (FRII-Q), and radio non-detected quasars within the ultraviolet (UV) Eigenvector 1 (E1) plane defined by the Mg II line width (\FWHMmg) and the UV Fe II strength (\rfeUV). This represents the first systematic comparison of BZQ blazars with their proposed parent population in the UV E1 framework. To minimize selection biases, we construct statistically representative control subsamples and perform matched-subsample analyses in redshift and Mg II luminosity. Population differences are assessed using one- and two-dimensional non-parametric tests, including the Kolmogorov–Smirnov, Anderson–Darling, Mann–Whitney U, and Fasano–Franceschini statistics.
We find that the three AGN populations occupy statistically distinct regions of the UV E1 plane. The dominant separation is driven by \FWHMmg, with BZQ blazars systematically exhibiting narrower line profiles than FRII-Q quasars, while radio non-detected quasars lie between both populations. In contrast, \rfeUV\ provides only limited discrimination between the RL subclasses, although it separates radio-loud from radio non-detected quasars at the population level. 
The observed ordering demonstrates that the UV E1 plane preserves the physical sequence expected from RL AGN unification. Inclination-dependent projection provides a natural first-order explanation for the Mg II line-width ordering. Additional differences in intrinsic BLR structure or accretion properties may contribute to the observed distributions, although the present analysis does not allow these effects to be separated independently from orientation.
\end{abstract}




\section{Introduction} 

Blazars constitute a subclass of radio-loud (RL) active galactic nuclei (AGN) characterized by relativistic jets closely aligned with the observer’s line of sight \citep{antonucci93,urrypadovani95}. This orientation leads to Doppler boosting and relativistic amplification of the emitted radiation \citep{sher68}, producing strong variability across the entire electromagnetic spectrum on timescales ranging from minutes to years \citep{guptaetal17,fanetal18}. Their spectral energy distributions (SEDs) are dominated by non-thermal jet emission, with synchrotron radiation extending from radio to UV/X-ray energies and inverse Compton (IC) emission dominating at higher energies.

Blazars are conventionally divided into three classes \citep[e.g.,][]{haroldetal21}, which represent different observational manifestations of relativistic jets:
(a) BZB objects, which exhibit featureless optical spectra or only weak narrow absorption/emission lines (typical BL Lac sources);
(b) BZQ objects, also known as flat-spectrum radio quasars (FSRQs), these sources display prominent broad emission lines; and
(c) BZG objects, in which the optical–UV continuum is dominated by host-galaxy emission while retaining blazar-like properties at other wavelengths.
{BZQs are unique in providing direct access to the broad-line region (BLR) through standard spectroscopic diagnostics.}

In this work we focus on BZQ-type blazars (FSRQs) and investigate their spectroscopic properties using the full width at half-maximum (FWHM) of the low-ionization broad \mgii\ emission line and the strength of the adjacent \feii\ UV complex. Because \mgiionly\ remains observable in the optical for intermediate-redshift quasars, it provides one of the most widely used virial diagnostics beyond the redshift range where \hb\ is accessible. \mgiionly\ is emitted in a region located farther from the central black hole than the \hb -emitting BLR, consistent with its systematically narrower width \citep{shenetal08,popovicetal19}.

To interpret the observed line widths and Fe II emission strengths, we place our analysis within the framework of the Eigenvector 1 (E1) parameter space \citep{byg92, sulentic00a}. E1 links key spectroscopic observables—broad-line widths, optical Fe II strength, and high-ionization line shifts—into what is commonly referred to as the quasar Main Sequence (MS), a cornerstone for understanding AGN diversity. {The optical E1 plane, defined by FWHM(\hb) versus \rfe = EW(\feii)/EW(\hb), reveals a continuous distribution traditionally divided into Population A and Population B. Population A (Pop. A) sources are characterized by FWHM(\hb) $\lesssim$ 4000 \kms, strong \feii\ emission, and frequently exhibit \civ\ blueshifts. In contrast, Population B (Pop. B) sources show broader \hb\ profiles, weaker \feii\ emission, and typically lack strong outflow signatures \citep{marzianietal01, zamfiretal08}, thereby occupying a well-defined region of the E1 plane characterized by broad low-ionization lines and comparatively weak \feii\ emission.} The Population A/B distinctions observed in the optical can be reproduced in the UV using the \mgii\ emission line together with the \feiiUV\ emission complex \citep[e.g.,][]{mengitsueetal23}, thereby enabling the extension of the E1 formalism to higher redshifts where \hb\ is no longer accessible. 

The E1 framework also highlights systematic differences between radio-loud (RL) and radio-quiet (RQ) quasars. Classical lobe-dominated RL quasars populate a restricted region of E1 space, predominantly within Pop. B \citep{sulenticetal03, zamfiretal08}. Within the RL population, unified models propose that Fanaroff–Riley type II \citep[FRII;][]{fr74} quasars represent the parent population of core-dominated RL sources. The latter correspond to systems viewed within $\approx 25^{\circ}$ of the jet axis \citep{urrypadovani95}, where Doppler boosting enhances the radio and optical continuum emission and produces rapid variability and apparent superluminal motion \citep{ghisellinietal93}.

Despite these advances, important gaps remain. Optical E1 studies have shown that RL/RQ quasars occupy different regions of the parameter space, consistent with the standard orientation-based unification models \citep{sulenticetal03,zamfiretal08}. \citet{gancietal19} later confirmed, using a radio-selected sample, that RL sources preferentially occupy Pop.~B and that radio loudness affects the \hb\ and \oiii\ profiles along the MS. Nevertheless, these studies were based almost exclusively on \hb, leaving the UV E1 plane largely unexplored. {Although the optical E1 sequence has become one of the principal frameworks for understanding AGN diversity, its applicability to relativistically beamed radio-loud AGN has remained largely unexplored. Determining whether BZQ blazars and their parent FRII quasars preserve the same physical ordering in the UV plane represents an important test of whether the E1 framework extends naturally to jetted AGN. Since the available samples originate from different surveys and selection strategies, we also evaluate the impact of selection effects through matched-subsample analyses.}

This provides the motivation for the present work. The main goal of this paper is to determine how BZQ-type blazars, FRII quasars and radio non-detected quasars are distributed across the UV E1 plane, as traced by \FWHMmg\ and \rfeUV\, and to interpret any similarities or differences within this framework. In Section \ref{sec:data_samples} we describe the data samples and selection criteria. Section \ref{sec:analysis} outlines the spectral analysis and statistical methodology. The results are presented in Section \ref{sec:results} and discussed in Section \ref{sec:discussion}. Our main conclusions are summarized in Section \ref{sec:conclusions}.

\section{Data samples} \label{sec:data_samples}

In this study we compare BZQ-type blazars with two control populations selected from the SDSS quasar catalogue. Together these three datasets allow us to investigate the location of jetted and non-jetted AGN within the UV E1 parameter space. The primary sample consists of flat-spectrum radio quasars (FSRQs; hereafter BZQs) selected from the fifth edition of the Roma-BZCAT catalogue \citep{massaroetal15}. The control populations are: (i) a radio-loud (RL) population of quasars with FRII radio morphology identified by \cite{guptaetal16}, and (ii) a population of radio non-detected quasars, representing a radio-quiet comparison sample \citep[][hereafter S11]{shenetal11}.

\subsection{FSRQ-type blazar sample}

Our spectroscopic blazar sample is drawn from the fifth edition of the Roma-BZCAT Multifrequency Catalogue of Blazars \citep[5BZCAT;][]{massaroetal15} and cross-matched with optical spectra from the sixteenth data release of the Sloan Digital Sky Survey (SDSS DR16; \citealt{ahumadaetal20}).
The spectra cover the wavelength range 3650–10400 \AA. To ensure inclusion of the \mgii\ line and the adjacent 3000 \AA\ continuum, we restricted the sample to sources with 0.34 $\leq z \leq$ 2.41, yielding 439 FSRQ-type blazars with \mgiionly\ coverage. This sample forms the basis of the spectral analysis presented in \cite{patinoetal26}. Sources were further classified into confirmed blazars (BZQ-C; N = 119), showing clear flaring activity; possible blazars (BZQ-P; N = 242), characterized by low-amplitude variability and compact or one-sided radio morphology; and non-confirmed sources (BZQ-nC; N = 78), lacking flaring behavior and exhibiting two-sided kiloparsec-scale radio morphology. Details of the classification procedure are given in \citealt{patinoetal26} and \citealt{guerrerogonzalezetal26}.

\subsection{FRII Radio-Loud Quasar Sample}

We selected quasars with FRII radio morphology from \cite{guptaetal16} to construct the radio-loud comparison sample. {Their catalog includes 797 FRII quasars spanning the redshift range $0.1 \leq z \leq 2.0$. To ensure homogeneous spectral coverage relative to the BZQ sample, we restricted the redshift range to $0.34 \leq z \leq 2.00$, guaranteeing that \mgiionly\ is fully covered in the SDSS spectra.} 


The SDSS DR7 quasar sample is nearly complete to \textbf{\textit{i}} $\approx$ 19.1 \citep{schneideretal10}, we derived the equivalent redshift-dependent g-band completeness limit following \cite{richardsetal01}. We adopted the conservative criterion $g \leq 19$ which minimizes spectroscopic incompleteness while ensuring adequate signal-to-noise ratios for reliable spectral measurements. 

To minimize contamination by strongly beamed sources, we analyzed optical variability using light curves from the Zwicky Transient Facility \citep[ZTF;][]{mascietal19}, Data Release 24, and it provides an eight-year photometric baseline with sufficient precision for objects brighter than $g \approx 19$, allowing a robust characterization of optical variability.
We restricted the analysis to sources with adequate cadence (at least one observation every two weeks on average) to ensure reliable timescale measurements. For each light curve and photometric band, we defined a quiescent reference level following \citet{meyeretal19}, representing the baseline flux outside major variability episodes. Candidate peaks above this level were identified, and their temporal evolution was characterized by fitting a linear trend to the logarithm of the flux as a function of time. The resulting logarithmic slope defines a characteristic timescale for a fractional flux increase.

Motivated by observational studies showing that blazars can display flux increases by a factor of about four on timescales of roughly 80 days \citep[e.g.][]{gauretal12,hayashidaetal15}, we classify a flare-like event as one with a statistically significant fit ($p<0.05$), a positive slope and a characteristic timescale of $\leq$ 80 days \citep{guerrerogonzalezetal26}. Sources where at least one flare was found in any of the two optical bands were excluded from the FRII subsample.  Only four objects satisfied these conditions and were removed, leaving a sample consistent with the variability properties expected for unbeamed FRII quasars (gradual, low-amplitude variations over timescales of weeks to months).

We further excluded sources lacking the required spectral measurements or exhibiting relative parameter uncertainties greater than 50 per cent. After the redshift selection, magnitude cut, variability filtering and quality requirements, the final radio-loud comparison sample comprises 308 FRII quasars, hereafter referred to as the FRII-Q sample.

\subsection{Control Sample of Radio Non-Detected Quasars}

To construct a control sample for comparison with the BZQ blazars, we extracted \mgii\ line properties, 3000  \AA\ continuum luminosities, \feiiUV\ and \mgiionly\ equivalent widths from the SDSS quasar catalogue of S11. We restricted the redshift range to $0.34 \leq z \leq 2.00$, the same as BZQ and FRII-Q samples.

{We excluded all sources detected in the Faint Images of the Radio Sky at Twenty-Centimeters survey \citep[FIRST;][]{beckeretal95} or NRAO VLA Sky Survey \citep[NVSS;][]{condonetal98} radio surveys, and removed any object for which upper limits were reported. This conservative approach ensures a purely radio non-detected sample and minimizes contamination from radio-loud or weakly radio-emitting sources. After these cuts, 78,302 objects remained.}

We then excluded sources lacking reliable measurements of (i) \FWHMmg , (ii) a fitted broad Mg II component, or (iii) \feiiUV\ equivalent width. Objects with relative uncertainties exceeding 50 per cent in any of these quantities were also discarded. The final control sample comprises 65,378 quasars, hereafter referred to as QSOs.

\section{Analysis} \label{sec:analysis}

\subsection{Spectral parameter determination} \label{ssec:spec_params}

The spectral parameters and associated uncertainties for the radio non-detected AGN and the FRII radio-loud samples were extracted from the SDSS quasar catalogue of S11. For the BZQ sample, spectra were shifted to the rest frame and corrected for Galactic extinction using the extinction law of \citet{cardellietal89}. Spectral fitting was performed with the Image Reduction and Analysis Facility \citep[IRAF;][]{tody86} using the {\tt specfit} task \citep{kriss94}, following the procedure described in \cite{patinoetal26}.

To ensure a homogeneous spectral decomposition across the BZQ sample, we limited the fit to the 2500–3100 \AA\ region (rest-frame) around \mgiionly , where the continuum is adequately represented by a single power law and prominent \feii\ emission \citep[][Fig. 1]{patinoetal26}. We assessed whether this restricted fitting interval introduces systematic differences by comparing our measurements with those reported by S11 for sources in common (see Fig. \ref{fig:comp_s11vsbzq}).
The fits included the \feiiUV\ template of \citet{vyw01}. 

The unresolved \mgiionly\ profile was represented by one Gaussian or, when required by the residuals, by the sum of a relatively narrow and a broad Gaussian. These components provide only a phenomenological description of the total profile and are not associated with distinct emitting regions. {Although Mg II profiles can often be represented by Lorentzian or multi-component functions \citep[e.g.][]{popovicetal19}, our purpose is to obtain the FWHM of the total reconstructed profile rather than to assign physical meaning to individual components. Its uncertainty reflects both the intrinsic random error due to spectral dispersion and the additional uncertainty introduced by the Gaussian decomposition of the line profile \cite[see ][for details]{amadorportesetal25}.}

The \mgiionly\ equivalent width (EW) was computed by integrating the line profile relative to the fitted continuum. The \feiiUV\ EW is defined following S11 over 2200–3090\AA. As most BZQ spectra do not extend to 2200\AA , the best-fit power-law continuum was extrapolated to 2200\AA , assuming the same fitted power-law shape. The full empirical \feii\ template \citep{vyw01} was scaled to the fitted \feiiUV\ emission, to have the same integrated flux within the fitted region. Then, the EW was calculated over 2200–3090 \AA\ using the scaled template and extrapolated continuum. Uncertainties in both \mgiionly\ and \feiiUV\ EWs were estimated following \citet[][Sec. 2.2.2]{{patinoetal26}}.

To verify the consistency of our measurements with previous SDSS analyses, we compared our results with those reported by S11. We cross-matched the BZQ sample with the SDSS quasar catalogue, obtaining 181 sources in common. 

\subsection{Statistical tests} \label{ssec:stat_tests}

To quantitatively compare the spectral properties of the three AGN samples, we apply non-parametric statistical tests to the distributions of \FWHMmg\ and \rfeUV. Because at least two of the samples (QSO and FRII-Q) exhibit departures from normality, we employ:

\begin{itemize}
\item the Kolmogorov–Smirnov (K-S) test \citep{KS51},
\item the two-sample Anderson–Darling (A-D) test \citep{AD54}, and
\item the Mann–Whitney U (MWU) test \citep{Hollander14}.
\end{itemize}

The K-S test is sensitive to differences in both the location and overall shape of the distributions. The A-D test extends this approach by placing greater weight on discrepancies in the distribution tails, which is particularly relevant for the skewed \FWHMmg\ and \rfeUV\ distributions observed in the UV E1 plane. The MWU test provides a complementary, rank-based evaluation of differences in central tendency without assuming a specific functional form for the distributions.
We adopt the A-D test as the primary non-parametric statistic. The K-S and MWU tests serve as consistency checks to verify the robustness of the results.

The test statistics and their assumptions are provided in Appendix \ref{app:stat_tests}. The results of these statistical comparisons are presented in the following Section, where they are used to assess the significance of the trends observed in the UV E1 plane.

\subsection{Radio Non-Detected Quasars: construction of representative subsamples}
\label{ssec:shen_subsample}

A key methodological challenge in this study arises from the large disparity in sample sizes among the three AGN populations (BZQ: 439; FRII-Q: 308; QSO: 65,378). Although non-parametric two-sample tests are generally robust, extreme differences in sample size can substantially increase statistical power, potentially yielding statistically significant but physically negligible differences. Constructing representative QSO subsamples therefore provides a balanced statistical framework for comparing the radio-loud and radio non-detected populations.

{Following the bootstrap methodology developed by \citet{patinoetal26}, we generated candidate QSO subsamples with sizes ranging from $n=100$ to $n=1000$. For each candidate size, 10,000 random realizations were produced and evaluated using five complementary similarity metrics: (i) the maximum ECDF difference, (ii) the Wasserstein distance, (iii) the standardized mean difference, (iv) the standard deviation ratio, and (v) differences in higher-order moments. For every realization and metric, the subsample providing the closest match to the parent QSO distribution was recorded. The results from all realizations and metrics were then combined to identify the subsample size that most consistently reproduces the one-dimensional statistical properties of the full QSO population.}

{This analysis identifies $n=600$ as the optimal subsample size. Independent representative QSO subsamples were therefore constructed for the \FWHMmg\ and \rfeUV\ analyses, each preserving the one-dimensional distribution of the corresponding parameter while maintaining a sample size comparable to those of the radio-loud populations. These representative subsamples are adopted throughout the one-dimensional statistical analyses presented below. Details of the bootstrap size-selection procedure are given in Appendix~\ref{app:QSO_subsample}.}

\subsection{Two-dimensional comparison of the UV E1 plane}
\label{ssec:2D}

{To complement the one-dimensional comparisons, we performed a direct statistical analysis of the joint UV E1 plane defined by \FWHMmg\ and \rfeUV. Unlike the representative subsamples described in the previous subsection, which preserve the marginal distribution of each parameter independently, the two-dimensional analysis requires preserving the joint structure of the UV E1 plane.}

{Bootstrap realizations with candidate sizes ranging from $n=100$ to $n=1000$ were therefore evaluated using five complementary multivariate similarity metrics: (i) the average one-dimensional K--S statistic, (ii) the average Wasserstein distance, (iii) the Mahalanobis distance between the two-dimensional means (multivariate Cohen's $d$), (iv) the logarithmic ratio of the covariance-matrix determinants, which measures the relative volumetric spread of the distributions, and (v) the average absolute difference in skewness and kurtosis. As in the one-dimensional analysis, 10,000 bootstrap realizations were generated for each candidate size. A subsample size of $n=600$ was adopted because it provides the best compromise between reproducing the joint distribution of the parent QSO population and maintaining a sample size comparable to those of the radio-loud samples.}

{The resulting representative QSO subsamples were compared with the BZQ and FRII-Q samples using the two-dimensional Fasano--Franceschini (F--F) test \footnote{https://github.com/wmpg/fasano-franceschini-test} \citep{fasanofranceschini87}, implemented through the algorithm of \citet{chowbrown25}. The F--F test extends the K--S statistic to two dimensions by comparing the cumulative distributions over the full parameter space. Statistical significance was estimated from 10,000 random permutations for each comparison, yielding a minimum resolved probability of $p=10^{-4}$. The corresponding F--F statistic is denoted by $D_{\rm FF}$ throughout this work. The resulting statistics are presented in Section~\ref{sec:results} together with the one-dimensional comparisons.}

\subsection{Matched-subsample construction}

{To evaluate whether the observed differences between the samples could be driven by selection effects, we constructed matched subsamples using a one-to-one nearest-neighbour algorithm without replacement. Since the parent catalogs exhibit different redshift and luminosity distributions, BZQ-C objects were matched independently to the QSO and FRII-Q samples using redshift and \mgiionly\ luminosity as control variables. The matching tolerances were chosen to maximize the number of matched pairs while ensuring statistically indistinguishable distributions in the control variables. After matching, the statistical comparisons were repeated for the remaining physical parameters using the statistical tests described in Section \ref{ssec:stat_tests}.}

\subsection{Black-hole masses and Eddington ratios}

{To investigate whether the observed differences in the UV E1 plane may be related to global accretion properties, we estimated black-hole masses, bolometric luminosities, and Eddington ratios for the three samples.}

{For the BZQ sample, the black-hole masses were estimated using three independent single-epoch virial calibrations based on the \mgiionly\ emission line: \citet{mcluredunlop04,kongetal06,shenliu12},}

\begin{equation}
\log\left(\frac{M_{\rm BH}}{M_\odot}\right) = \log A + B \log \left( \frac{L}{L_0} \right) + 2 \log \left( \frac{{\rm FWHM(Mg\,II)}}{{\rm km\,s^{-1}}} \right),
\end{equation}

{where the coefficients $(A,B,L_0)$ depend on the adopted calibration. Rather than selecting a single estimator, we adopted as the fiducial black-hole mass the mean of the three logarithmic estimates. The minimum and maximum estimates among the three calibrations were retained as an indicator of the systematic dispersion introduced by the choice of virial relation. For the FRII-Q and QSO samples, previously published virial black-hole masses available in the original catalogs were adopted. Because the available catalog masses and the BZQ estimates do not use an identical virial prescription, absolute offsets between the distributions may include calibration-dependent systematics. We therefore use these quantities primarily as comparative accretion proxies and avoid interpreting their absolute normalization.}

{No single Mg II virial calibration has been demonstrated to be universally superior over the luminosity and redshift ranges spanned by our samples. We therefore adopt the average of the three widely used calibrations in order to reduce the dependence of the analysis on any individual prescription, while the dispersion among them provides an estimate of the systematic uncertainty associated with the choice of calibration.}

{Bolometric luminosities were derived from the continuum luminosity at 3000~\AA. For the BZQ sample, the disk continuum was estimated from the \mgiionly\ line luminosity using Eq. (11) of \cite{patinoetal26}. This line luminosity was used because the observed 3000-\AA\ continuum may contain a non-thermal jet contribution. For the QSO and FRII-Q samples, the published \mgiionly\ luminosities were converted directly into the corresponding continuum luminosities. A luminosity-dependent bolometric correction was then applied following the prescription of \cite{netzer19}. The Eddington luminosity was computed as $L_{Edd} = 4\pi G M_{BH}m_{pc} / \sigma_T$ (erg s$^{-1}$) and the Eddington ratio was calculated as $L_{bol}/L_{Edd}$. The uncertainties in the logarithmic quantities were propagated from the published measurement errors of luminosity and FWHM whenever available. However, these formal uncertainties do not include the intrinsic systematic scatter of single-epoch virial mass estimators, typically of the order of $\sim0.3$--$0.5$ dex, nor the uncertainty associated with the adopted bolometric correction.}

{Finally, it is important to emphasize that virial black-hole masses are not independent of the \mgiionly\ line width because all calibrations explicitly include \FWHMmg\ as a squared term. Consequently, if the line width is affected by orientation, part of the inferred black-hole mass may also retain an orientation-dependent component. For this reason, the derived black-hole masses and Eddington ratios are interpreted only as comparative
accretion proxies rather than absolute estimates.}

\section{Results} \label{sec:results}

\subsection{Consistency with the S11 measurements}
\label{ssec:s11_comparison}

\begin{figure*}
\includegraphics[width=0.33\textwidth]{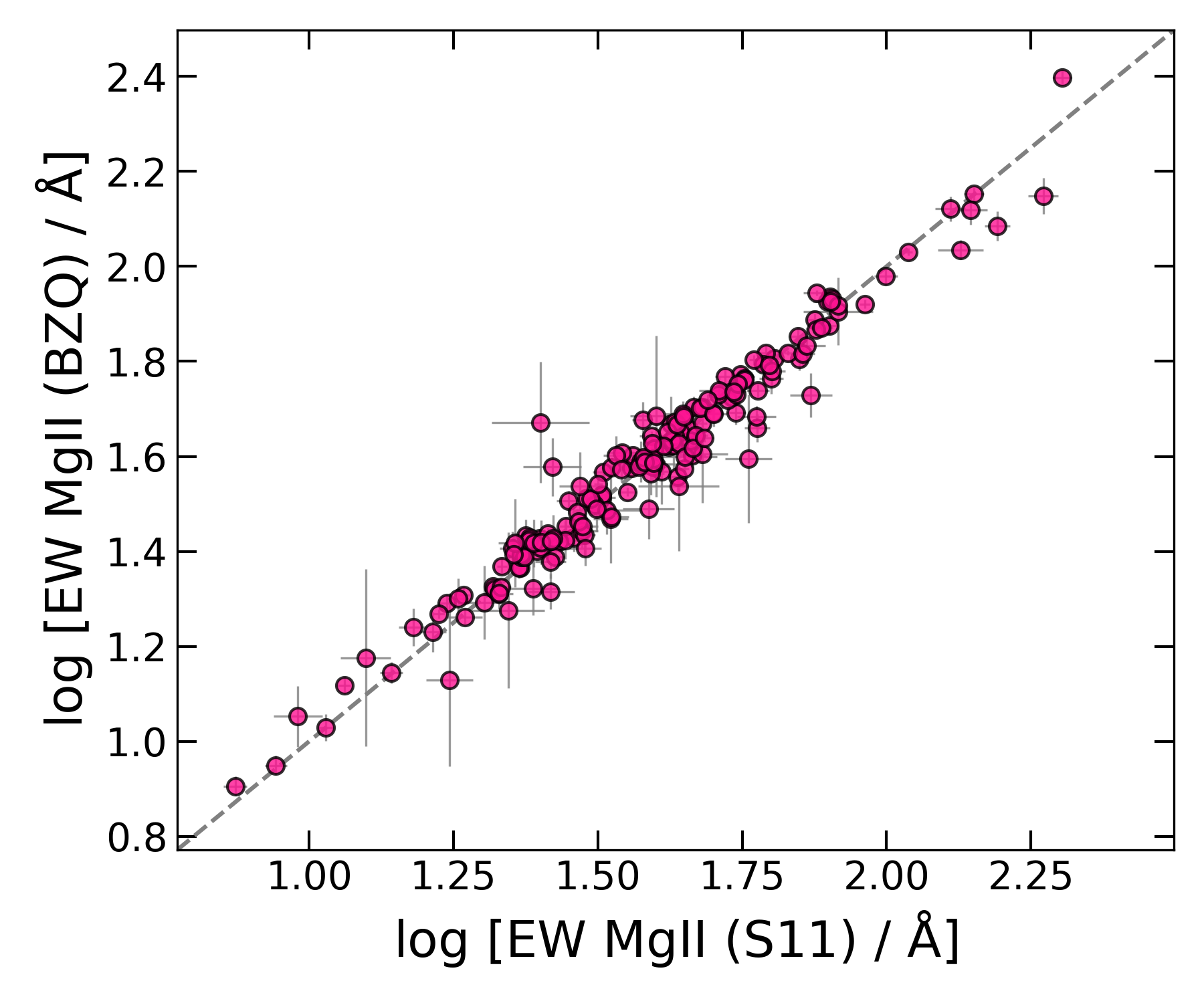}
\includegraphics[width=0.33\textwidth]{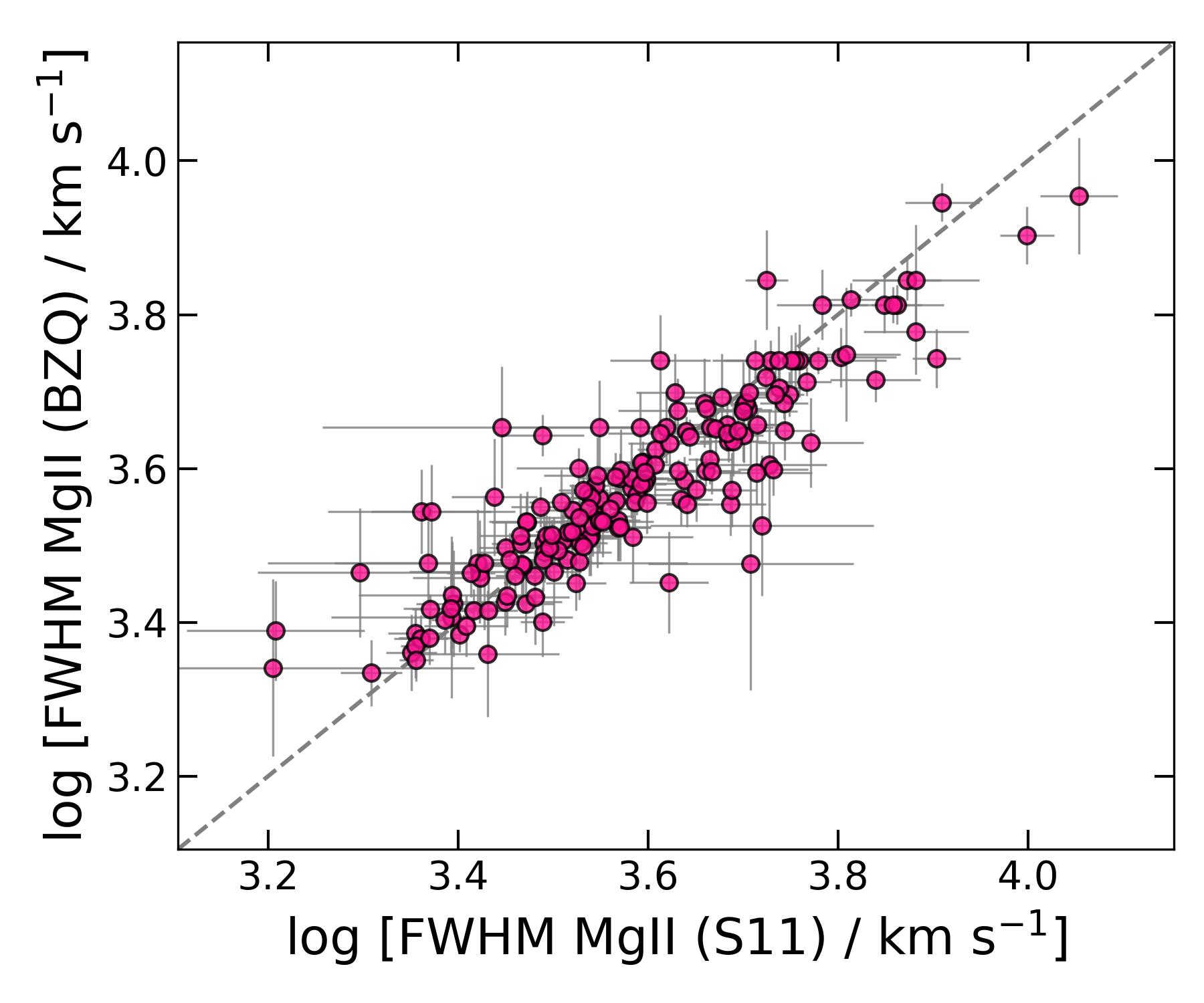}
\includegraphics[width=0.33\textwidth]{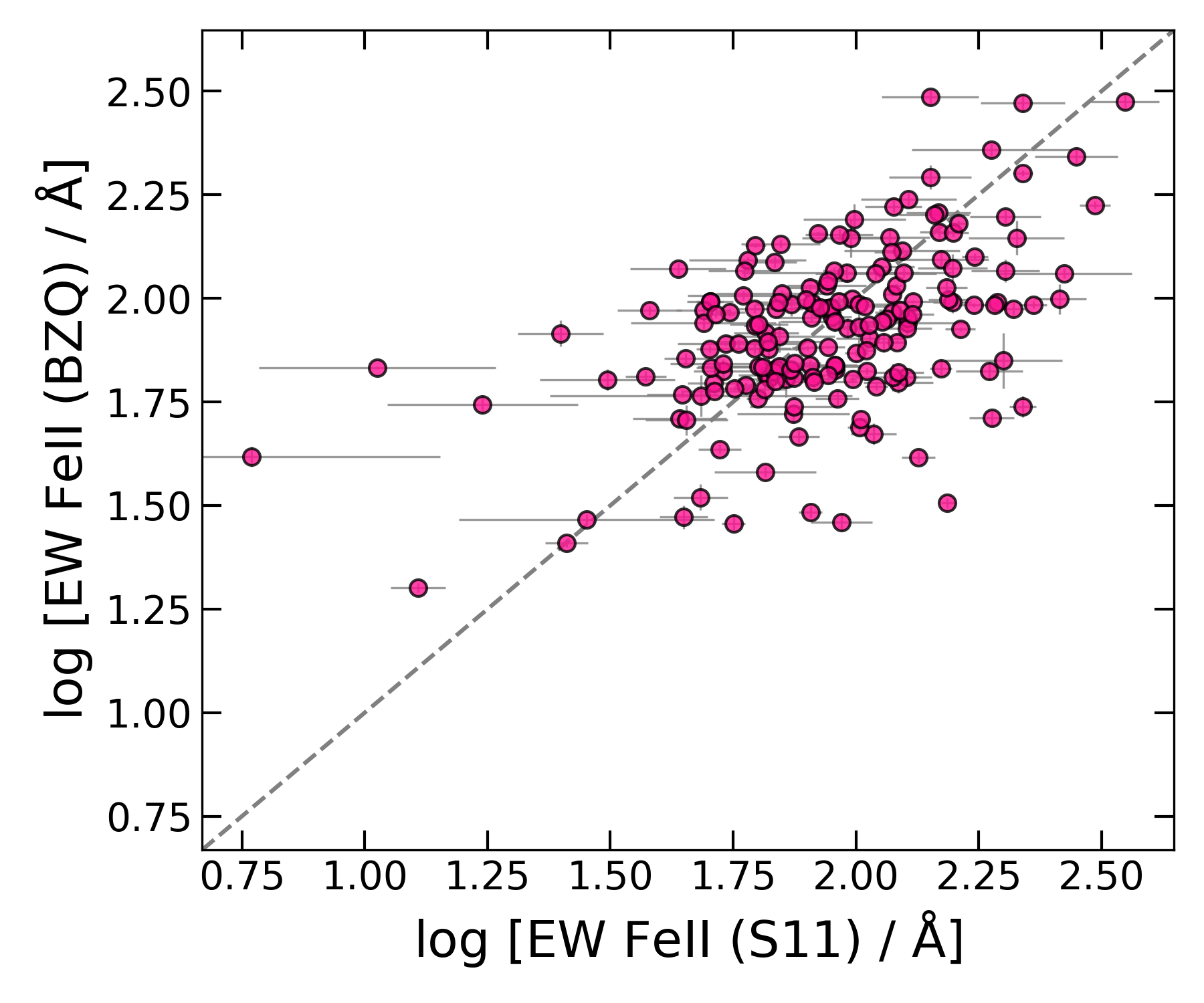}
\caption{Comparison of our measurements with S11 for sources in common: EW of \mgiionly\ (\textit{left}), \FWHMmg\ (\textit{middle}), and EW of \feiiUV\ (\textit{right}). The dashed line indicates the one-to-one relation.}
\label{fig:comp_s11vsbzq}
\end{figure*}

\begin{figure}
\includegraphics[scale=0.5]{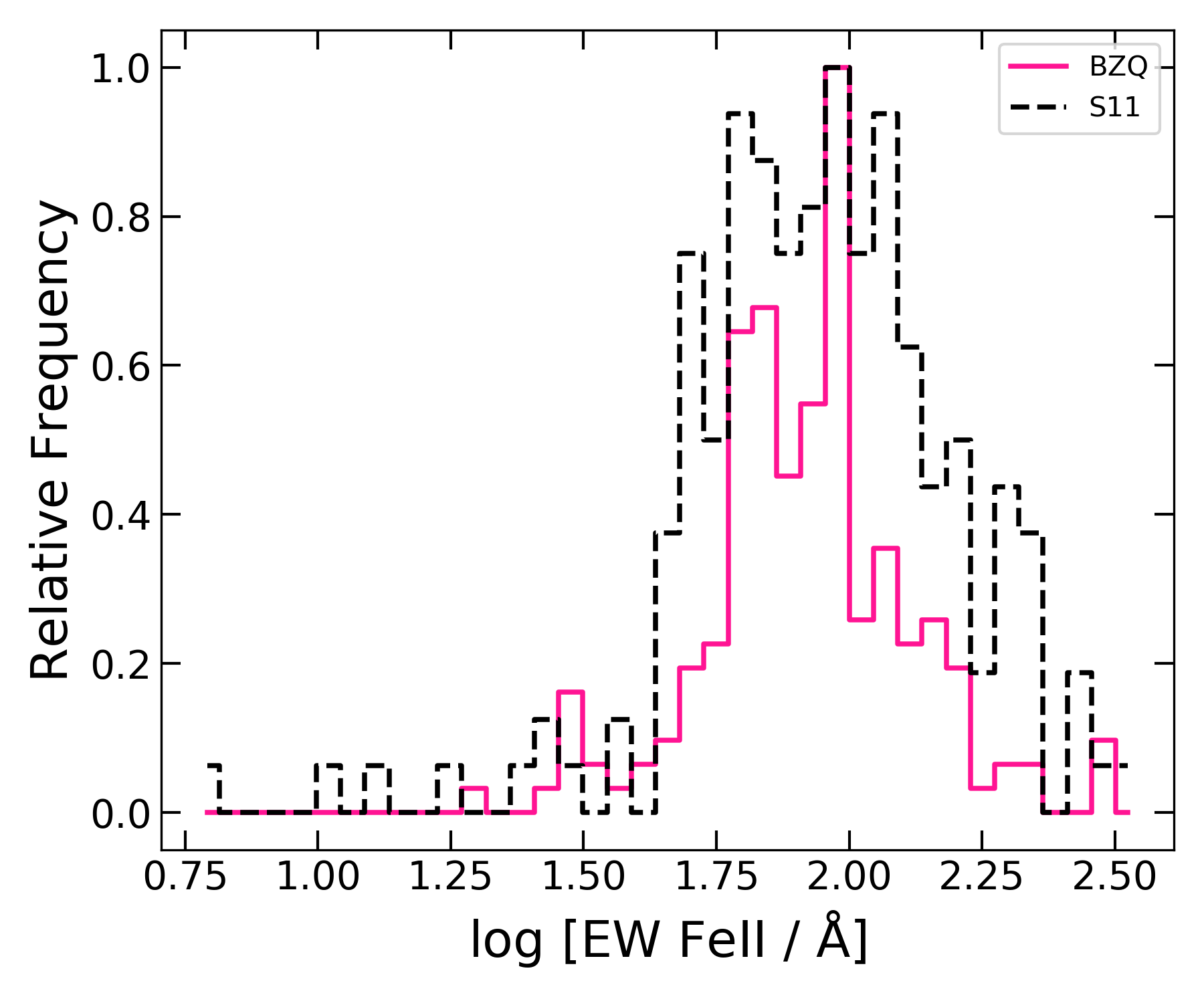}
\includegraphics[scale=0.5]{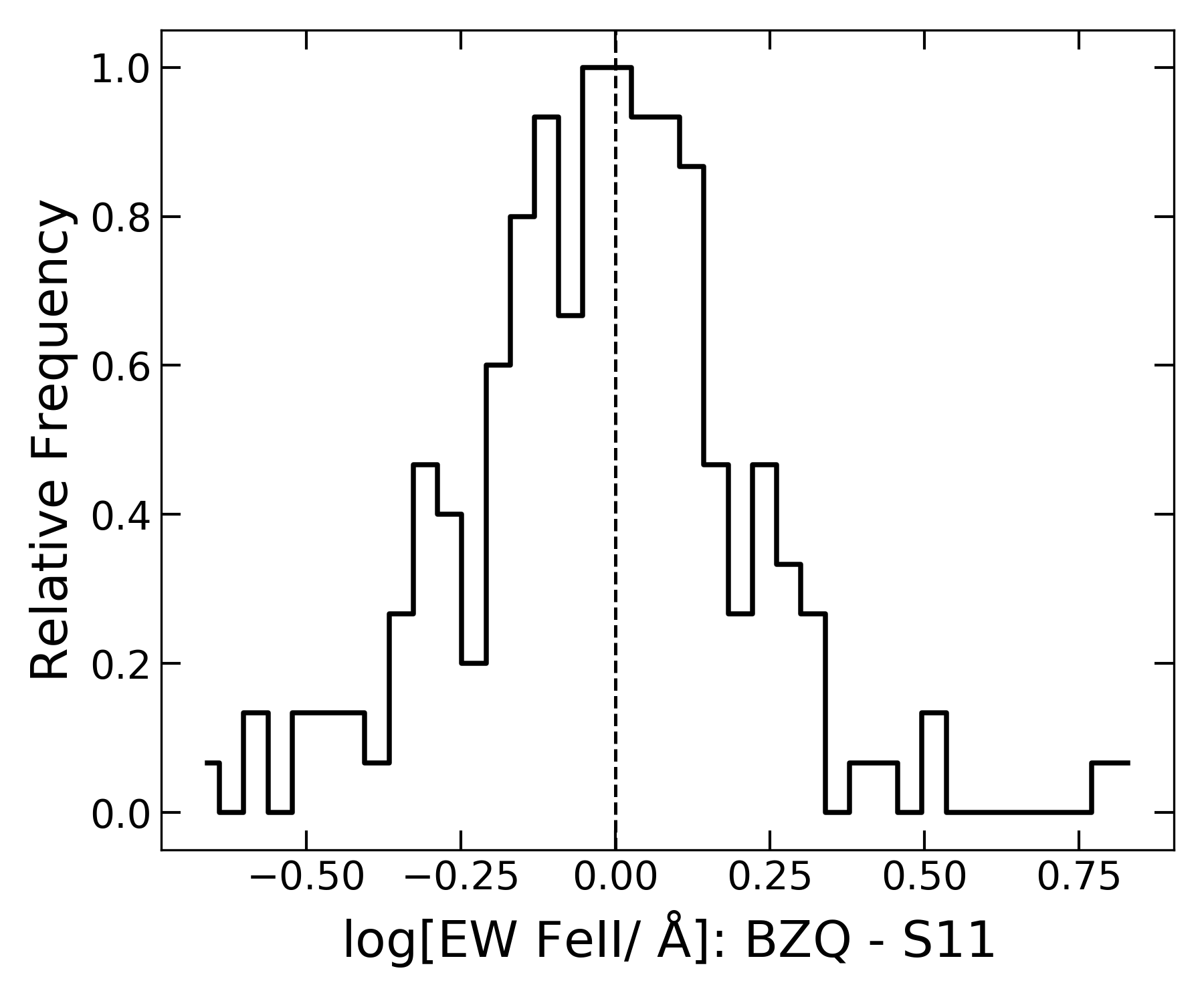}
\caption{\textit{Upper:} histogram of EW \feiiUV\ for S11 (dashed) and BZQ (filled). \textit{Lower:} distribution of differences between the two measurement methods.}
\label{fig:histograms_comp}
\end{figure}

{Figure \ref{fig:comp_s11vsbzq} compares our measurements with those of S11 for the 181 sources in common. The mean logarithmic differences, defined as ours minus S11, are 
$\Delta \log {\rm EW}(\feii)=-0.021\pm0.227$, 
$\Delta \log {\rm EW}(\mgiionly)=0.002\pm0.050$, and 
$\Delta \log FWHM_{MgII}=-0.007\pm0.065$.}

{These small offsets are consistent with zero within the quoted dispersions (as seen in Fig. \ref{fig:histograms_comp}), indicating no strong global systematic shift; thus, our fitting method yields spectral parameters fully compatible with those of S11. However, the substantially larger scatter in EW(Fe II), $\sigma \approx$0.23 dex, reflects the greater sensitivity of this measurement to continuum placement, Fe II-template scaling, wavelength coverage, and extrapolation toward 2200\AA.}

\subsection{Main sequence in the UV plane} \label{ssec:UVplane}

\begin{figure*}
\includegraphics[width=\textwidth]{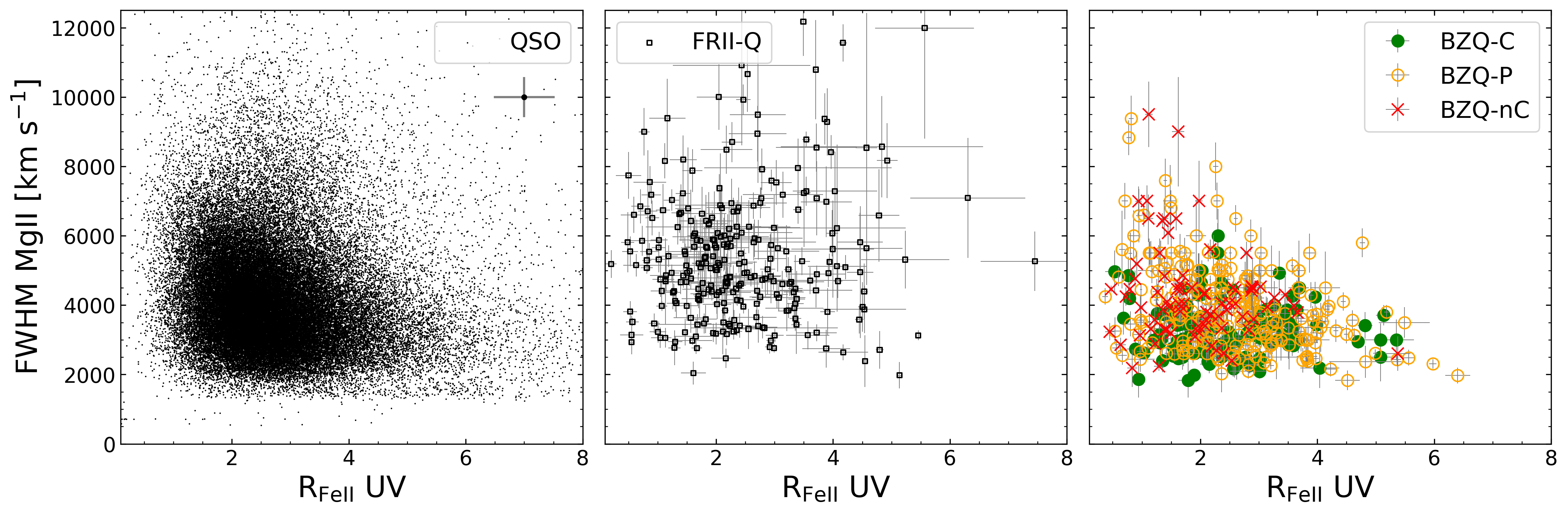}
\caption{Distribution of the three AGN samples (BZQ, FRII-Q, QSO) in the UV E1 plane defined by \rfeUV\ vs.\ \FWHMmg . \textit{Left}: radio non-detected QSOs from S11. \textit{Center}: radio-loud AGN with FRII morphology (FRII-Q). \textit{Right}: BZQ blazars. In the left panel, points represent individual QSOs and the cross is the typical uncertainty as the interquartile range (IQR) of the uncertainty distributions for each parameter. In the right panel, symbols indicate classification: BZQ-C (green circles), BZQ-P (yellow open circles), and BZQ-nC (red crosses).}
\label{fig:E1_samples}
\end{figure*}

Figure \ref{fig:E1_samples} shows the distribution of the three AGN samples in the UV E1 plane defined by \rfeUV\ versus \FWHMmg\ diagram. The QSO sample spans the full extent of the UV MS (\textit{left panel}), serving as a reference distribution. The cross in the upper right corner indicates the representative uncertainty. The FRII-Q population is preferentially distributed toward larger \FWHMmg\ values and moderate-to-low \rfeUV\ (\textit{center panel}), broadly corresponding to the region associated with Pop. B in optical E1 studies. In contrast, the BZQ sample (\textit{right panel}), including the BZQ-C, BZQ-P, and BZQ-nC subclasses, occupies a partially overlapping region but extends toward systematically smaller \FWHMmg\ values.

These qualitative differences already suggest that the three populations occupy different regions of the UV E1 plane. However, visual inspection alone cannot establish whether the observed offsets are statistically significant, motivating the quantitative analyses presented in the following sections.

\subsection{Selection effects}

\begin{figure}
\includegraphics[scale=0.49]{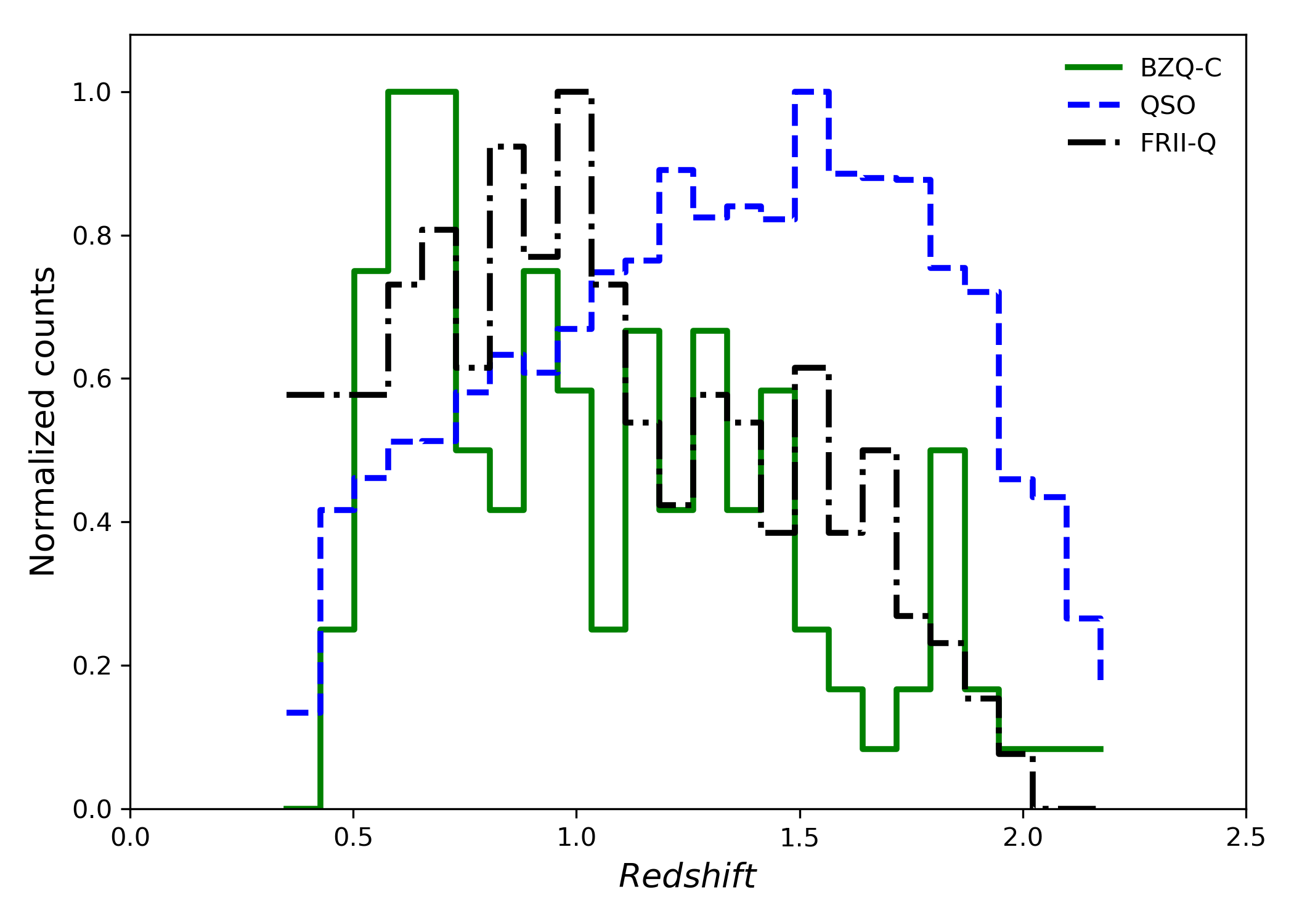}
\includegraphics[scale=0.49]{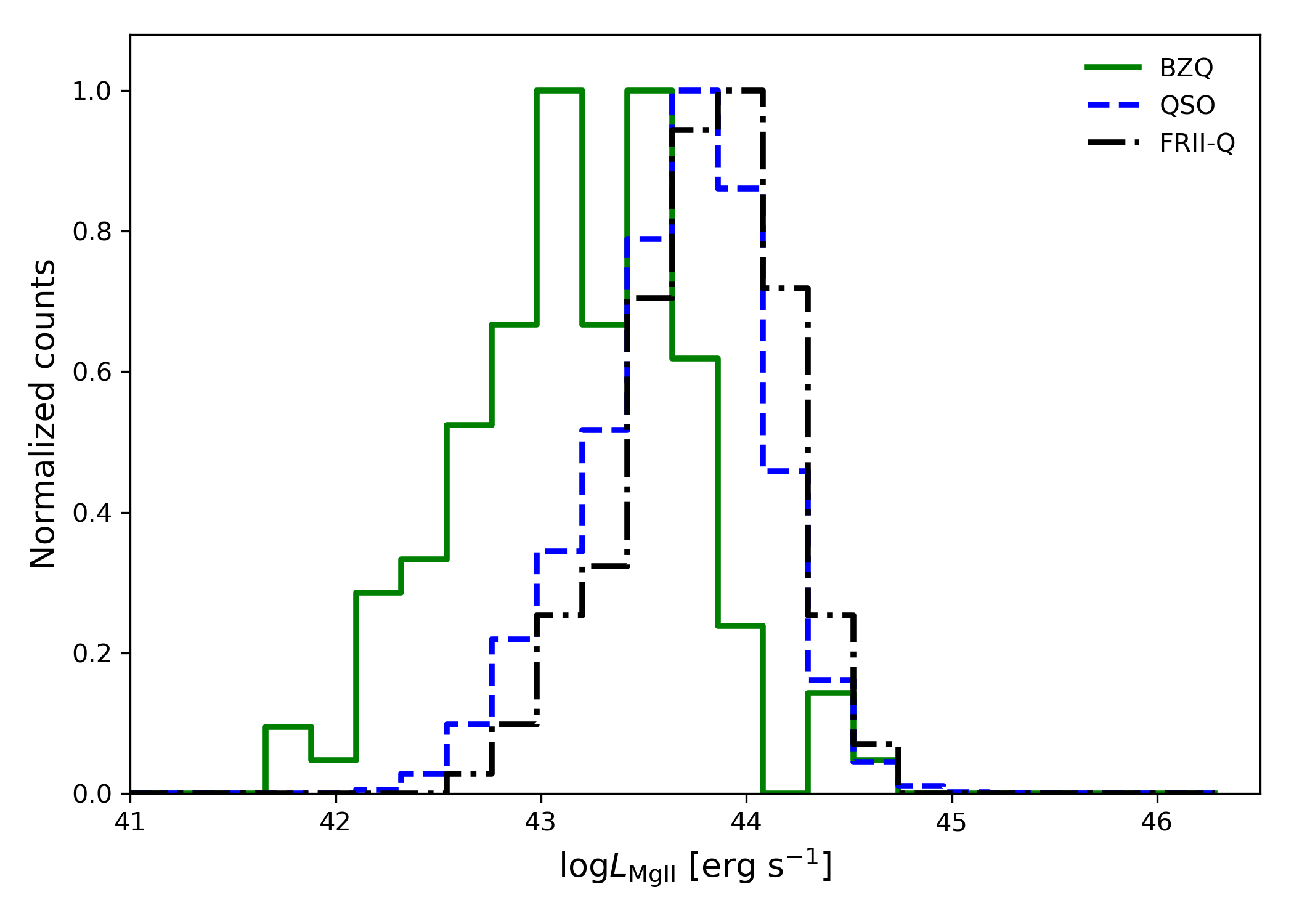}
\caption{{Histograms of the redshift (\textit{upper}) and \mgiionly\ luminosity (\textit{lower}) distributions for the three samples: BZQ-C (green), FRII-Q (black) and QSO (blue).}}
\label{fig:hist_z_LMg}
\end{figure}

{Figure~\ref{fig:hist_z_LMg} compares the redshift and \mgiionly\ line luminosity distributions of the BZQ-C, FRII-Q, and QSO samples. The three populations substantially overlap in both parameters, although their distributions are not identical. The BZQ sample has the highest mean redshift ($\langle z\rangle=1.41$; IQR $=0.84$--$1.88$), followed by the QSO sample ($\langle z\rangle=1.33$; IQR $=0.97$--$1.69$), whereas the FRII-Q population is shifted toward lower redshift ($\langle z\rangle=1.03$; IQR $=0.70$--$1.35$). Likewise, the \mgiionly\ luminosity distributions show substantial overlap, with the FRII-Q and QSO samples extending to slightly higher luminosities than the BZQs. The broad overlap in both redshift and line luminosity indicates that the three samples probe comparable regions of parameter space, enabling meaningful comparisons of their emission-line properties.}

{Although the overlap is considerable, the distributions are not identical. These systematic offsets motivate the matched-subsample analysis presented below, which explicitly controls for redshift and \mgiionly\ luminosity before comparing the UV E1 properties of the different AGN populations.}

\subsection{Black-hole masses and Eddington ratios}

\begin{figure}
\includegraphics[scale=0.49]{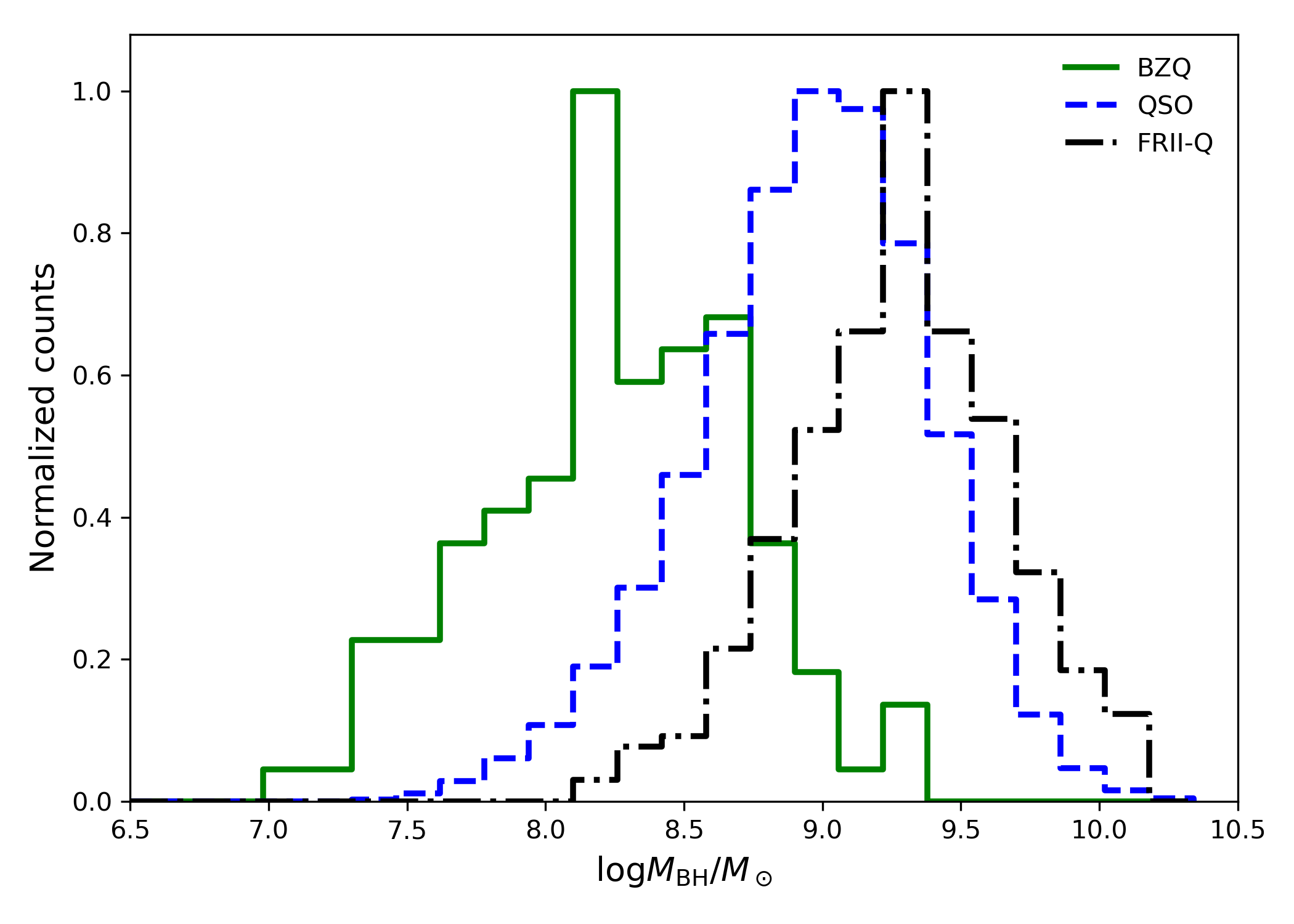}
\includegraphics[scale=0.49]{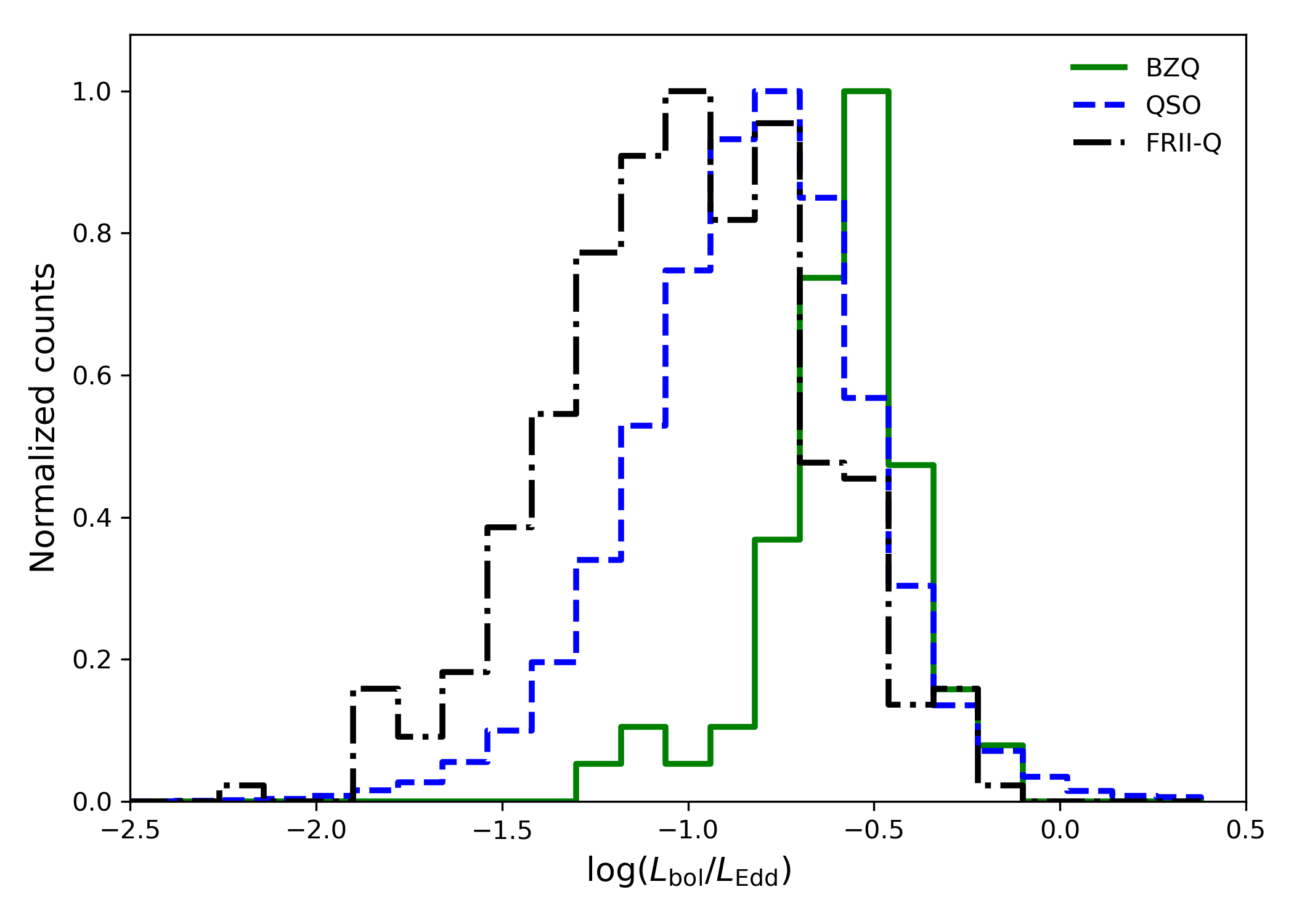}
\caption{{Histograms of the black hole mass ($M_{BH}$, \textit{upper}) and Eddington ratio ($L_{bol}/L_{Edd}$, \textit{lower}) for the three samples. Color code is the same as Fig. \ref{fig:hist_z_LMg}.}}
\label{fig:hist_mbh_redd}
\end{figure}

{Figure~\ref{fig:hist_mbh_redd} compares the distributions of black-hole mass and Eddington ratio for the three samples. FRII-Q objects preferentially populate the highest black-hole masses, whereas BZQ-C sources are shifted toward higher Eddington ratios. The QSO population occupies an intermediate position in both parameters. These trends are consistent with the expectation that the radio-loud subclasses differ not only in their emission-line properties, but also in their characteristic accretion states. Because virial black-hole masses are derived from \mgiionly\ line widths, they cannot be interpreted independently of possible orientation effects. Nevertheless, the observed distributions provide useful context for interpreting the UV E1 trends discussed in Section~\ref{sec:discussion}.}

\subsection{Equivalent widths of \mgiionly\ and \feiiUV}
\label{ssec:ew_comparison}

\begin{figure*}
    \centering
    \includegraphics[scale=0.48]{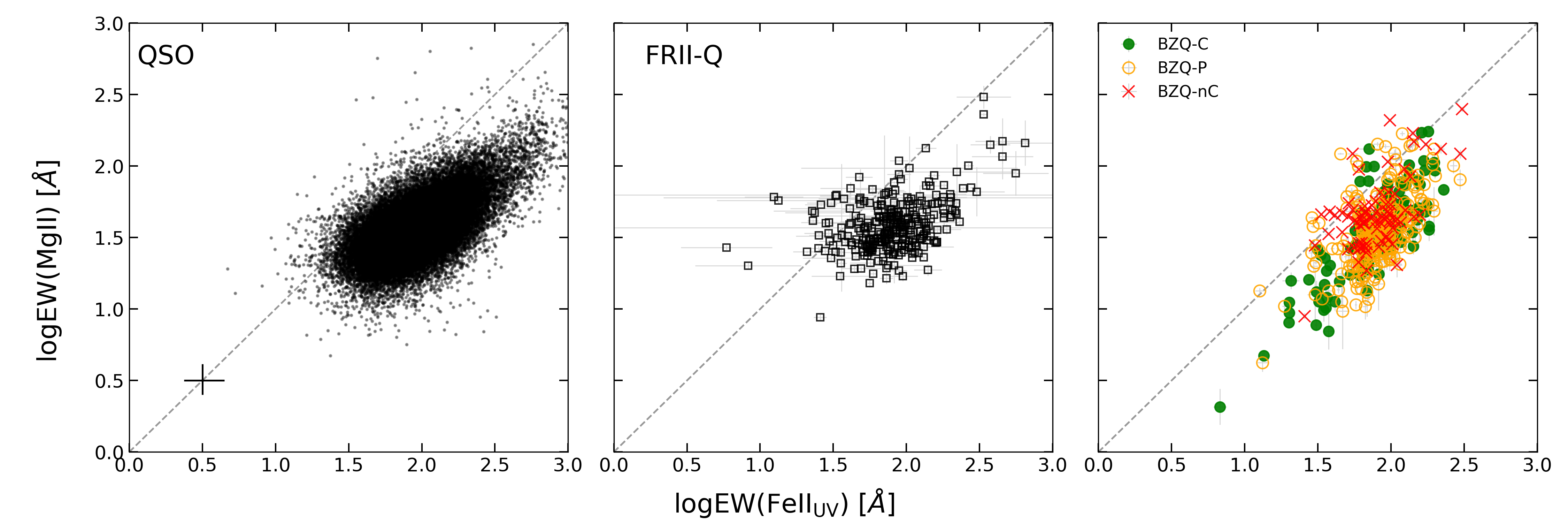}
    \caption{{Distribution of the three AGN samples (BZQ, FRII-Q, and QSO) in the $\log$ EW(\mgiionly) versus $\log$ EW(\feiiUV) plane. The dashed grey line indicates equality. The color coding is the same as in Fig.~\ref{fig:E1_samples}.}}
    \label{fig:ew_mgii_feii}
\end{figure*}

{Figure~\ref{fig:ew_mgii_feii} shows a strong positive correlation between the equivalent widths of \mgiionly\ and the \feiiUV\ complex for all AGN populations. Pearson correlation coefficients are $r=0.645$ (QSO), $0.431$ (FRII-Q), $0.804$ (BZQ-C), $0.571$ (BZQ-P), and $0.581$ (BZQ-nC), all with $p<10^{-7}$. Orthogonal distance regression yields slopes are of order unity: $0.959\pm0.003$ (QSO), $1.166\pm0.086$ (BZQ-C), $1.314\pm0.087$ (BZQ-P), and $1.392\pm0.161$ (BZQ-nC). Although the BZQ subclasses exhibit slightly steeper slopes, all populations follow the same overall sequence.} 

{The existence of this common EW(\mgiionly)--EW(\feiiUV) relation indicates that the population-level differences discussed throughout this work are not driven by the absolute strength of either emission feature alone. }

\subsection{ECDFs and test statistics (K-S, A-D, and MWU)} \label{ssec:ECDFs}

\begin{figure*}
    \centering
    \includegraphics[scale=0.5]{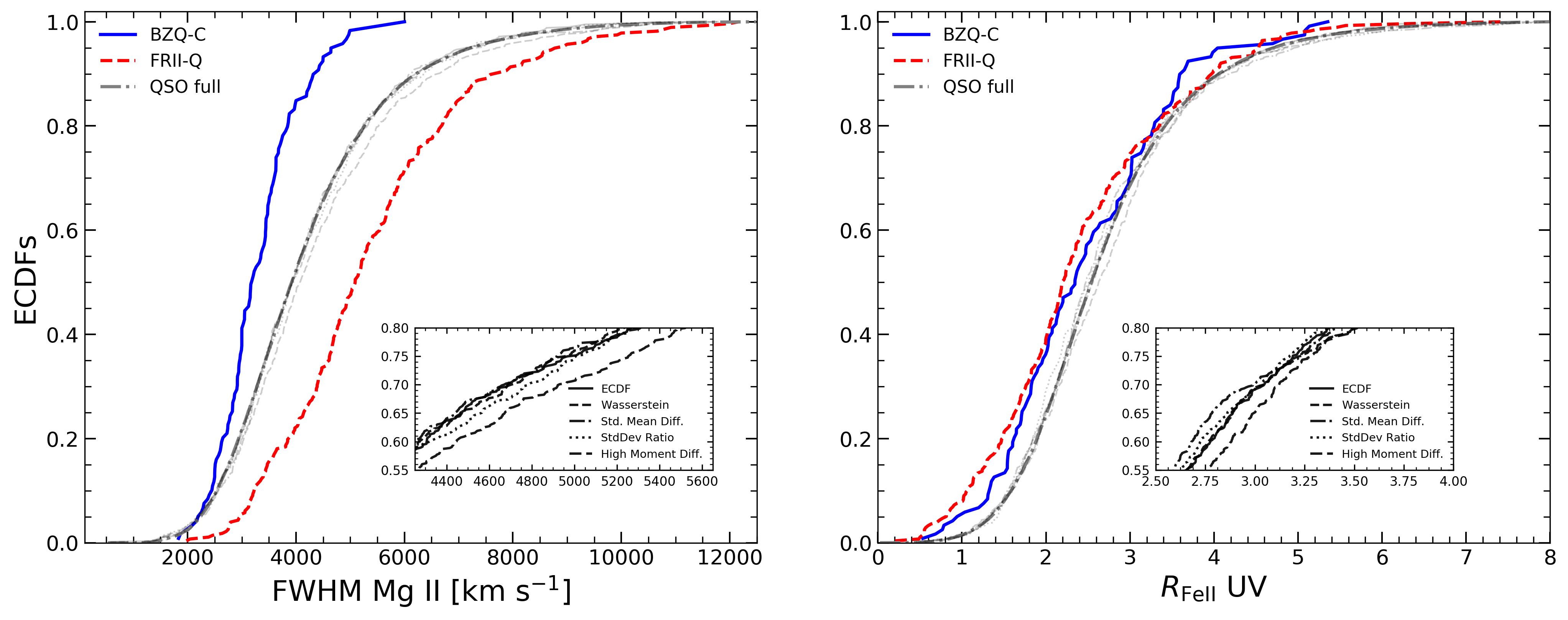}
    \caption{{Empirical cumulative distribution functions (ECDFs) of \FWHMmg\ (\textit{left}) and \rfeUV\ (\textit{right}) for the BZQ-C, FRII-Q, and full QSO samples. Insets show zoomed regions ($4250 \le \mathrm{FWHM} \le 5650 \mathrm{km s^{-1}}$ and $0.55 \le \mathrm{ECDF} \le 0.8$, and $2.5 \le \mathrm{R_{FeII}\ UV} \le 4.0$ and $0.55 \le \mathrm{ECDF} \le 0.8$ in ECDF for \rfeUV) to illustrate subtle differences among the five parameter-specific constructed QSO subsamples of $n\approx600$.}}
    \label{fig:ecdf_inset}
\end{figure*}

\subsubsection{\FWHMmg}

{Figure~\ref{fig:ecdf_inset} (left panel) shows the ECDFs of \FWHMmg\ for the BZQ-C, FRII-Q, and QSO samples. Both radio-loud populations are systematically offset from the radio non-detected QSOs, with the largest separation occurring in the direct BZQ-C--FRII-Q comparison.}

{These differences are consistently confirmed by the K-S, A-D, and MWU tests (Table~\ref{tab:ecdf_tests_matrix}) for all five statistically representative QSO subsamples, demonstrating that the results are independent of the adopted subsampling criterion.
The FRII-Q sample yields systematically larger A-D statistics ($A^2\approx42$--61) than BZQ-C ($A^2\approx27$--36), indicating a stronger departure from the QSO distribution. The direct BZQ-C--FRII-Q comparison gives the largest test statistics ($D_{\rm KS}=0.620$, $A^2=85.83$), confirming a strong separation between the two radio-loud subclasses. The relatively stable K-S statistics among the five representative QSO subsamples indicate that the inferred population differences are insensitive to the criterion used to construct the reference sample.}

\subsubsection{\rfeUV}

{Independent representative QSO subsamples were constructed for \FWHMmg\ and \rfeUV\ in order to reproduce the corresponding ECDFs of the parent QSO population (Fig.~\ref{fig:ecdf_inset}, right panel).}

{Compared with \FWHMmg, the \rfeUV\ distributions show substantially weaker statistical separation (Table~\ref{tab:rfeuv_tests_matrix}). All three tests indicate significant differences between the radio-loud samples and the QSO reference subsamples, with FRII-Q exhibiting the largest deviations ($D_{\rm KS}=0.18$--0.21, $A^2=22$--30). By contrast, the direct BZQ-C--FRII-Q comparison yields only modest statistics ($D_{\rm KS}=0.133$, $A^2=2.14$), consistent with the strong overlap of both radio-loud subclasses in \rfeUV .The lower K-S and A-D statistics reflect a substantially larger overlap among the distributions than observed for FWHM.}

\begin{table*}
\centering
\caption{{Results of the Kolmogorov--Smirnov (K-S), Anderson--Darling (A-D), and Mann--Whitney U (MWU) tests comparing the \FWHMmg\ distributions of the radio-loud samples with the five statistically representative QSO subsample of $n\approx600$.s.}}
\label{tab:ecdf_tests_matrix}
\begin{tabular}{lccccc}
\hline
\textbf{Statistic} & \textbf{ECDF} & \textbf{Wasserstein} & \textbf{Std. Mean Diff.} & \textbf{StdDev Ratio} & \textbf{High Moment Diff.} \\
\midrule

\multicolumn{6}{l}{\textbf{BZQ full sample}}\\
K-S $D$ & 0.137 & 0.146 & 0.143 & 0.163 & 0.192 \\
A-D $A^2$ & 12.29 & 12.73 & 12.62 & 15.24 & 24.60 \\
MWU $U$ & $1.12\times10^{5}$ & $1.12\times10^{5}$ & $1.11\times10^{5}$ & $1.10\times10^{5}$ & $1.04\times10^{5}$ \\
\midrule

\multicolumn{6}{l}{\textbf{BZQ-C}}\\
K-S $D$ & 0.336 & 0.338 & 0.342 & 0.352 & 0.382 \\
A-D $A^2$ & 26.93 & 27.25 & 27.98 & 28.94 & 35.72 \\
MWU $U$ & $2.25\times10^{4}$ & $2.24\times10^{4}$ & $2.22\times10^{4}$ & $2.21\times10^{4}$ & $2.05\times10^{4}$ \\
\midrule

\multicolumn{6}{l}{\textbf{FRII-Q}}\\
K-S $D$ & 0.334 & 0.329 & 0.336 & 0.313 & 0.284 \\
A-D $A^2$ & 60.91 & 59.86 & 60.25 & 56.34 & 41.72 \\
MWU $U$ & $1.31\times10^{5}$ & $1.30\times10^{5}$ & $1.31\times10^{5}$ & $1.29\times10^{5}$ & $1.24\times10^{5}$ \\
\midrule

\multicolumn{6}{l}{\textbf{BZQ-C vs FRII-Q (direct comparison)}}\\
K-S $D$ & 0.620 & -- & -- & -- & -- \\
A-D $A^2$ & 85.83 & -- & -- & -- & -- \\
MWU $U$ & 5552 & -- & -- & -- & -- \\
\bottomrule
\end{tabular}
\\
\textbf{Notes.}
The K-S statistic $D$ measures the maximum absolute difference between two ECDFs and is sensitive to global distributional differences.
The A-D statistic $A^2$ gives additional weight to deviations in the distribution tails.
The MWU statistic $U$ tests for differences in central tendency by assessing whether values from one sample tend to be systematically larger or smaller than those from the other.
\end{table*}

\begin{table*}
\centering
\caption{{Results of the Kolmogorov--Smirnov (K-S), Anderson--Darling (A-D), and Mann--Whitney U (MWU) tests comparing the \rfeUV\ distributions of the radio-loud samples with the five statistically representative QSO subsamples of $n\approx600$.}}
\label{tab:rfeuv_tests_matrix}
\begin{tabular}{lccccc}
\hline
\textbf{Statistic} & \textbf{ECDF} & \textbf{Wasserstein} & \textbf{Std. Mean Diff.} & \textbf{StdDev Ratio} & \textbf{High Moment Diff.} \\
\midrule

\multicolumn{6}{l}{\textbf{BZQ full sample}} \\
K-S $D$ & 0.165 & 0.162 & 0.148 & 0.152 & 0.174 \\
A-D $A^2$ & 17.00 & 18.00 & 15.15 & 14.93 & 22.56 \\
MWU $U$ & $1.08\times10^{5}$ & $1.08\times10^{5}$ & $1.11\times10^{5}$ & $1.11\times10^{5}$ & $1.04\times10^{5}$ \\
\midrule

\multicolumn{6}{l}{\textbf{BZQ-C}} \\
K-S $D$ & 0.139 & 0.139 & 0.132 & 0.120 & 0.152 \\
A-D $A^2$ & 3.05 & 3.31 & 2.48 & 2.13 & 4.81 \\
MWU $U$ & $3.10\times10^{4}$ & $3.10\times10^{4}$ & $3.17\times10^{4}$ & $3.19\times10^{4}$ & $2.99\times10^{4}$ \\
\midrule

\multicolumn{6}{l}{\textbf{FRII-Q}} \\
K-S $D$ & 0.197 & 0.191 & 0.181 & 0.184 & 0.206 \\
A-D $A^2$ & 24.93 & 25.86 & 23.13 & 22.54 & 29.66 \\
MWU $U$ & $7.49\times10^{4}$ & $7.49\times10^{4}$ & $7.63\times10^{4}$ & $7.70\times10^{4}$ & $7.25\times10^{4}$ \\
\midrule

\multicolumn{6}{l}{\textbf{BZQ-C vs FRII-Q (direct comparison)}} \\
K-S $D$ & 0.133 & -- & -- & -- & -- \\
A-D $A^2$ & 2.14 & -- & -- & -- & -- \\
MWU $U$ & $2.04\times10^{4}$ & -- & -- & -- & -- \\
\bottomrule
\end{tabular}
\end{table*}

\subsection{Two-dimensional distribution in the UV E1 plane}
\label{ssec:2D_results}

{The two-dimensional F-F tests reject the null
hypothesis that the QSO and radio-loud samples are drawn from the same joint distribution for all five QSO subsampling criteria. The QSO--BZQ comparison yields $D_{\rm FF}=3.06$--$3.35$, the QSO--BZQ-C comparison $D_{\rm FF}=3.38$--$3.51$, and the QSO--FRII-Q comparison $D_{\rm FF}=5.11$--$5.48$, with $p<10^{-4}$ in all cases.}

{The direct comparisons between the RL populations also reject the null hypothesis, yielding $D_{\rm FF}=6.07$ for BZQ--FRII-Q and $D_{\rm FF}=5.61$ for BZQ-C--FRII-Q, both with $p<10^{-4}$. Thus, the null hypothesis is rejected for every comparison when \FWHMmg\ and \rfeUV\ are considered jointly and are robust across the adopted QSO subsampling criteria.}

\subsection{Matched-subsample}

Table~\ref{tab:matched_results} summarizes the results of the
matched-subsample analysis using redshift and \mgiionly\ luminosity as control variables. We obtained 118 matched BZQ-C--QSO pairs and 44 matched BZQ-C--FRII-Q pairs. In both comparisons, the K-S and MWU tests show no significant differences in the control variables, confirming that the matching procedure produced subsamples with comparable redshift and \mgiionly\ luminosity distributions.

{For the BZQ-C--QSO matched samples, black-hole mass and Eddington ratio remain significantly different according to both tests. For \rfeUV, however, the results are test-dependent: the MWU test indicates a significant systematic shift in the ranked values ($p\sim10^{-4}$), whereas the K-S test does not reject the null hypothesis that the full distributions are consistent ($p\sim$0.1). The difference in \FWHMmg\ is not significant according to either test.}

The matched-subsample analysis represents the most stringent test performed in this work. By controlling simultaneously for redshift and \mgiionly\ luminosity, it demonstrates that the systematic separation between BZQ-C and FRII-Q cannot be explained solely by differences in sample selection. The remaining differences therefore cannot be explained solely by the two controlled selection variables and require further physical interpretation, as presented in Section~\ref{sec:discussion}.

\begin{table*}
\centering
\caption{{Results of the matched-subsample analysis using redshift and \mgiionly\ luminosity as control variables. The table reports the number of matched pairs, the median values of each parameter for the two samples, and the $p$-values from the K-S and MWU tests. The first two rows in each block correspond to the matching variables.}}
\label{tab:matched_results}
\begin{tabular}{llrrrrrr}
\hline
\textbf{Comparison} &
\textbf{Parameter} &
\textbf{$N_{\rm pair}$} &
\textbf{Median 1} &
\textbf{Median 2} &
\textbf{K-S $D$} &
\textbf{K-S $p$} &
\textbf{MWU $p$} \\
\midrule

\multicolumn{8}{l}{\textbf{BZQ-C vs QSO}}\\
$ $ & $z$ & 118 & 1.006 & 1.008 & 0.025 & 1.00 & 0.977 \\

$ $ & $\log\ L_{\rm MgII}$ & 118 & 43.190 & 43.191 & 0.017 & 1.00 & 0.983 \\

$ $ & $\log\ (M_{\rm BH}/M_{\odot})$ & 118 & 8.247 & 8.554 & 0.288 & $\sim 10^{-5}$ & $\sim 10^{-7}$ \\

$ $ & $\log\ (L_{\rm bol}/L_{\rm Edd})$ & 118 & -0.567 & -0.861 & 0.559 & $\sim 10^{-17}$ & $\sim 10^{-17}$ \\

$ $ & $\log\ {\rm FWHM_{MgII}}$ & 118 & 3.502 & 3.528 & 0.169 & $\sim 10^{-1}$ & 0.456 \\

$ $ & $\log\ {\rm R_{FeII,UV}}$ & 118 & 0.371 & 0.442 & 0.203 & $\sim 10^{-1}$ & $\sim 10^{-4}$ \\
\midrule

\multicolumn{8}{l}{\textbf{BZQ-C vs FRII-Q}}\\
$ $ & $z$ & 44 & 0.990 & 0.972 & 0.068 & 1.00 & 0.838 \\

$ $ & $\log\ L_{\rm MgII}$ & 44 & 43.591 & 43.598 & 0.159 & 0.595 & 0.512 \\

$ $ & $\log\ (M_{\rm BH}/M_{\odot})$ & 44 & 8.555 & 9.235 & 0.795 & $\sim 10^{-13}$ & $\sim 10^{-13}$ \\

$ $ & $\log\ (L_{\rm bol}/L_{\rm Edd})$ & 44 & -0.507 & -1.292 & 0.909 & $\sim 10^{-17}$ & $\sim 10^{-15}$ \\

$ $ & $\log\ {\rm FWHM_{MgII}}$ & 44 & 3.536 & 3.760 & 0.727 & $\sim 10^{-11}$ & $\sim 10^{-11}$ \\

$ $ & $\log\ {\rm R_{FeII,UV}}$ & 44 & 0.373 & 0.357 & 0.209 & 0.264 & 0.835 \\
\bottomrule
\end{tabular}
\\
\textit{Notes.} {Luminosities are expressed in erg s$^{-1}$, black-hole masses in solar masses, and FWHM in km s$^{-1}$. Parameters are reported in logarithmic units except for redshift. The BZQ-C sample is listed first in both comparisons. Values of $p<0.05$ indicate rejection of the null hypothesis at the nominal 5 per cent significance level.}
\end{table*}

\section{Discussion}
\label{sec:discussion}

{The statistical analyses reveal systematic differences among the three AGN populations in both the individual UV E1 parameters and their joint distribution. In this section we discuss the physical implications of these results, focusing first on the role of orientation in establishing the UV Eigenvector~1 sequence and then on the evidence that additional intrinsic BLR and accretion-related properties also contribute.}

\subsection{Orientation establishes the first-order UV E1 sequence}
\label{ssec:orientation}

\begin{figure}
\centering
\includegraphics[width=0.5\textwidth]{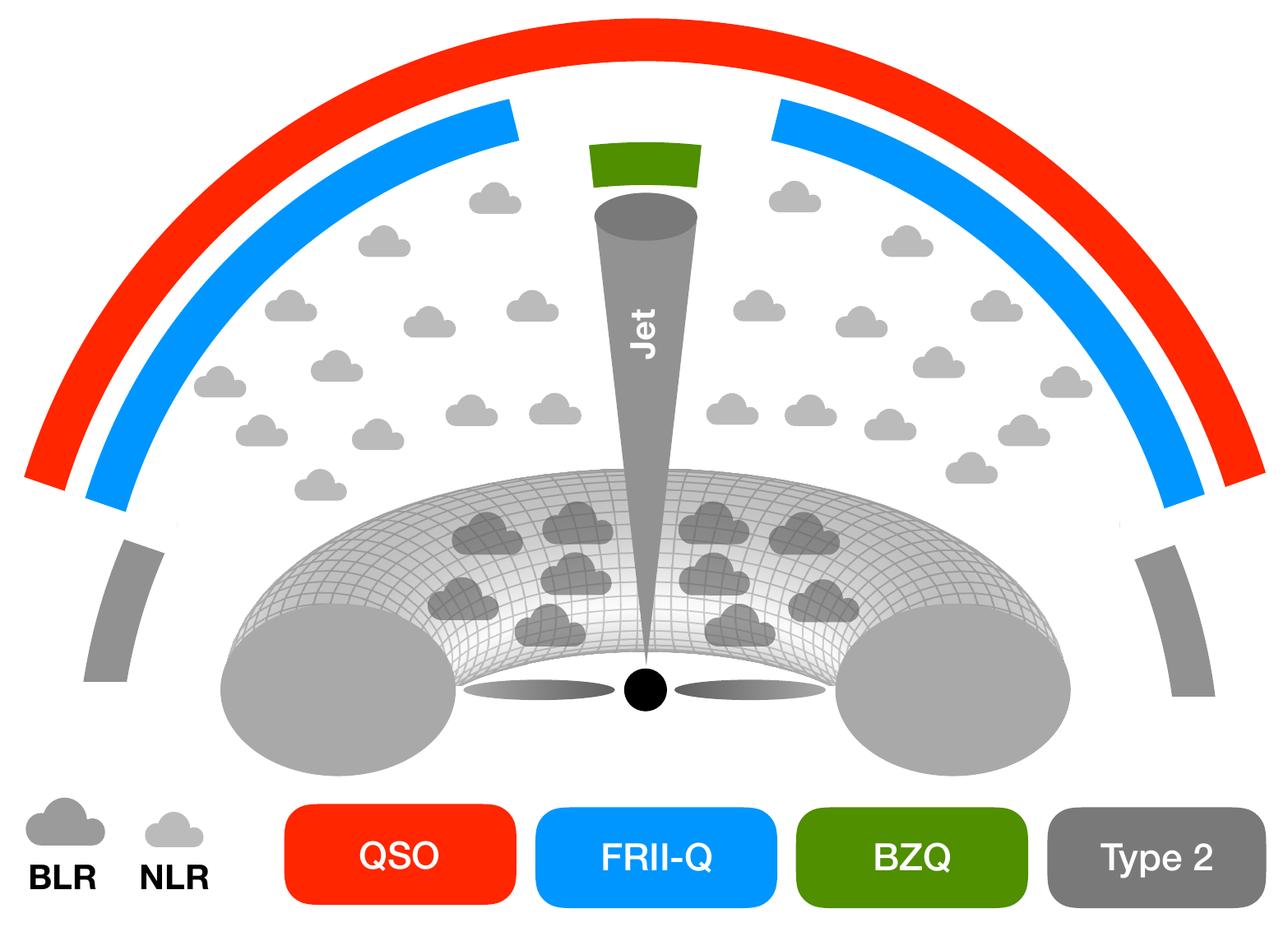}
\caption{Schematic, not to scale, of the orientation framework for the radio-loud AGN populations. BZQs (green) are observed close to the jet axis, whereas FRII-Q quasars (blue) are generally viewed at larger inclinations. The radio non-detected QSO sample (red) is shown as a broader comparison population spanning unobscured viewing angles. At the largest inclinations, the BLR becomes obscured and the source is observed as a Type~2 AGN (grey).}
\label{fig:sketch}
\end{figure}

{Orientation is a central component of radio-loud AGN unification models. If the BLR has a partially flattened or axisymmetric geometry, the projected line-of-sight velocity field is expected to depend on inclination \citep[e.g.,][]{wills86,antonucci93,urrypadovani95,decarlietal08}. Figure~\ref{fig:sketch} illustrates this framework: BZQs are observed close to the jet axis, whereas FRII-Q quasars are generally viewed at larger inclinations. The radio non-detected QSO sample is included as a broader comparison population spanning unobscured viewing angles.}

{Within this framework, the ordering of the \FWHMmg\ distributions follows the expectation of inclination-dependent projection. BZQ-C sources exhibit systematically narrower \mgiionly\ profiles than FRII-Q quasars, whereas the radio non-detected QSOs occupy an intermediate regime. If the BLR velocity field contains a significant rotational component, increasing inclination naturally produces larger projected velocities and consequently broader observed emission lines.}

{This interpretation is statistical and does not imply that \FWHMmg\ provides a direct inclination measurement for individual objects. Nevertheless, the observed ordering agrees remarkably well with the orientation sequence predicted by the radio-loud unification scenario.}

{The same interpretation is supported by the two-dimensional UV E1 analysis. The Fasano--Franceschini test shows that BZQs, FRII-Q quasars, and QSOs occupy statistically distinct regions of the joint \FWHMmg-\rfeUV\ plane, while the one-dimensional tests consistently identify the strongest population separation in \FWHMmg .}

{Taken together, these results indicate that orientation naturally explains the first-order organization of the UV Eigenvector~1 plane. This behavior is precisely what is expected within the standard AGN unification framework.}

\subsection{Orientation and possible additional physical effects}
\label{ssec:intrinsic}

Although orientation successfully explains the first-order ordering of the UV E1 sequence, it cannot by itself reproduce all of the observed population differences.

{The matched-subsample analysis provides an important test of whether the observed differences between BZQ-C and FRII-Q sources arise primarily from differences in the redshift and luminosity distributions of the parent samples. After matching in redshift and Mg II luminosity, the difference in \FWHMmg\ remains highly significant, while the two-dimensional F--F test continues to reject the null hypothesis that the two radio-loud populations occupy the same region of the UV E1 plane. Thus, the observed separation cannot be attributed solely to these selection variables.}

{Importantly, however, matching in redshift and Mg II luminosity does not control for orientation. The persistence of the \FWHMmg\ difference is therefore fully consistent with the inclination-dependent interpretation discussed in Section \ref{ssec:orientation}. If the Mg II-emitting BLR has a flattened velocity field, BZQ-C sources viewed closer to the jet axis are expected to display systematically narrower line profiles than FRII-Q quasars observed at larger inclinations.}

{The differences in virial black-hole mass and Eddington ratio found in the matched samples should likewise not be regarded as independent evidence for intrinsic physical differences. Mg II-based single-epoch mass estimators explicitly contain a ${\rm FWHM}^2$ dependence. For the matched BZQ-C–FRII-Q samples, the median difference of $\Delta\log {\rm FWHM}\simeq0.224$ dex alone corresponds to an approximately (0.45)-dex contribution to the inferred difference in $\log M_{\rm BH}$. Because $L_{\rm Edd}$ depends directly on black-hole mass, the derived Eddington ratios consequently retain the same orientation-dependent contribution.}

{Our results therefore support orientation as the primary explanation for the systematic Mg II line-width ordering between BZQ-C and FRII-Q sources. Additional differences in BLR geometry, kinematics, or accretion state may contribute to the residual dispersion and to the detailed structure of the UV E1 plane, but the present analysis does not allow these effects to be separated quantitatively from orientation. Establishing their relative importance will require independent black-hole mass or inclination estimates and an explicit model of the BLR geometry.}

The relativistic jet may also affect the continuum and emission-line profiles, but this influence is expected to be episodic. Non-thermal dominance \citep[NTD;][]{shawetal12} quantifies the jet contribution to the optical--UV continuum. During strong flares \citep[NTD $\gtrsim 2$;][]{patinoetal16}, the jet may temporarily outshine the accretion disk and produce measurable line variations \citep[e.g.,][]{leontavaresetal13,chavushyanetal20,amayaalmazanetal21,amayaalmazanetal22,amadorportesetal25}. However, only 11.8\% of the BZQ sample is classified as jet dominated \citep{patinoetal26}. Consequently, episodic jet activity is unlikely to determine the global UV E1 sequence identified in this work.

\subsection{\rfeUV\ in the UV main sequence}
\label{ssec:rfeUV}

{The behavior of \rfeUV\ differs markedly from that of \FWHMmg. Although the radio-loud populations differ from the radio non-detected QSOs, the BZQ-C and FRII-Q distributions overlap substantially. After matching in redshift and \mgiionly\ luminosity, their \rfeUV\ distributions become statistically indistinguishable, demonstrating that this parameter has considerably less discriminatory power than \FWHMmg.}

{The EW(\feiiUV)--EW(\mgiionly) relation provides additional context. All three AGN populations follow the same positive correlation, indicating that stronger Fe\,II UV emission is generally accompanied by stronger Mg\,II emission. Variations in \rfeUV\ therefore primarily reflect changes in the relative scaling of two closely related low-ionization emission components rather than large variations in either emission feature individually.}

{The physical interpretation of \rfeUV\ remains non-unique. Although the \feiiUV\/\mgiionly\ ratio has frequently been discussed as a metallicity indicator \citep[e.g.,][]{dietrichetal03,jiangetal24}, it is also affected by gas density, ionization parameter, radiative transfer, microturbulence, and the spectral shape of the ionizing continuum. In contrast to the optical \rfe\ parameter, \rfeUV\ compares two low-ionization species that likely originate in similar partially ionized regions of the BLR. Their similar response naturally reduces the dynamic range of the ratio and limits its ability to distinguish AGN subclasses.}

{The two-dimensional F--F analysis nevertheless demonstrates that the UV E1 plane separates the radio-loud populations when both parameters are considered jointly. Since the one-dimensional \rfeUV\ distributions overlap strongly whereas the \FWHMmg\ distributions remain clearly separated, the multivariate distinction is primarily driven by \mgiionly\ line width, with \rfeUV\ providing only a weaker secondary contribution.}

{The observed EW(\mgiionly)--EW(\feiiUV) relation is consistent with the SDSS analysis of \citet{calderoneetal17}, who reported a tight correlation between Mg\,II and UV Fe\,II emission. Likewise, \citet{sameshimaetal17} found that EW(\mgiionly) decreases with increasing Eddington ratio whereas EW(\feiiUV) exhibits a much weaker dependence. Together with the results of \citet{clowesetal16}, these studies suggest that the scatter in \rfeUV\ primarily reflects differences in BLR physical conditions and accretion state rather than abundance alone.}

Overall, our results indicate that the dominant population ordering within the UV Eigenvector~1 plane is established by \FWHMmg. Orientation provides the natural explanation for this first-order sequence. {Additional intrinsic BLR structure and accretion-related properties may contribute to the observed scatter and population differences, but their effects cannot be independently separated from orientation with the present analysis.} In contrast, \rfeUV\ acts primarily as a complementary parameter that traces the relative response of low-ionization BLR emission but has only limited discriminatory power between BZQ and FRII quasars.

\section{Conclusions}
\label{sec:conclusions}

We have investigated the distributions of BZQ blazars, FRII radio-loud quasars, and radio non-detected quasars within the UV Eigenvector~1 plane defined by \FWHMmg\ and \rfeUV. Combining one- and two-dimensional statistical tests with matched-subsample analyses, we reach the following main conclusions.

\begin{itemize}

\item {\textbf{The UV Eigenvector~1 plane distinguishes the three AGN populations.}
The one-dimensional non-parametric tests and the two-dimensional F--F analysis show that BZQs, FRII-Q quasars, and radio non-detected QSOs occupy statistically distinct regions of the UV E1 plane. These results remain robust across the different QSO subsampling criteria and cannot be attributed solely to the large imbalance in sample sizes. Matching in redshift and \mgiionly\ luminosity further demonstrates that the separation between BZQ-C and FRII-Q persists after controlling for the principal differences between the parent samples.}

\item {\textbf{\FWHMmg\ is the dominant population discriminator, whereas \rfeUV\ plays a secondary role.}
BZQ-C sources preferentially exhibit narrower \mgiionly\ profiles, FRII-Q quasars extend toward broader lines, and radio non-detected QSOs occupy an intermediate regime. In contrast, the BZQ-C and FRII-Q populations overlap substantially in \rfeUV, and their matched subsamples remain statistically consistent in this parameter. The common positive EW(\mgiionly)--EW(\feiiUV) relation followed by all populations further indicates that the two low-ionization emission components respond to related BLR conditions. Thus, the main population ordering in the UV E1 plane is driven by the \mgiionly\ line width rather than by \feiiUV\ strength.}

\item {\textbf{Orientation provides a natural first-order explanation for the observed sequence.}
{The ordering from narrow-lined BZQ-C sources to broad-lined FRII-Q quasars follows the expectation of inclination-dependent projection in a flattened BLR and is therefore consistent with radio-loud AGN unification. The differences in virial black-hole mass and Eddington ratio cannot be regarded as independent evidence for additional intrinsic differences because both quantities inherit the inclination dependence of \FWHMmg. Additional BLR geometry, kinematic, or accretion-related effects remain possible, but distinguishing them from orientation requires independent diagnostics or explicit orientation modeling. Episodic jet-dominated states are unlikely to determine the global trends because they characterize only a small fraction of the BZQ sample.}
}

\end{itemize}

More broadly, these results extend the physical interpretation of the Eigenvector~1 sequence to relativistically beamed AGN. They show that the UV plane provides a unified framework for investigating the interplay between jet orientation, broad-line region structure, and accretion in radio-loud quasars. {Future comparisons between the UV and optical Eigenvector 1 planes, together with independent orientation and accretion diagnostics, will help determine whether intrinsic physical parameters contribute to the spectroscopic diversity beyond the inclination-dependent trends identified here.}

\section*{Acknowledgements}

We thank the anonymous referee for providing constructive comments that helped to improve the manuscript. The work of T.M.B.-R. has been sponsored by Secretaría de Ciencia, Humanidades, Tecnología e Innovación (SECIHTI)-México through the program Estancias Posdoctorales por México. This work was supported by the Max Planck Institute for Radio Astronomy (MPIfR) - Mexico Max Planck Partner Group led by V.M.P.-A. J.U.G.-G. and M.M.-B. acknowledge support from SECIHTI-México program for Ph.D. and MsC students, respectively. C.A.N. acknowledges the support from grants SECIHTI CBF2023-2024-1418, PAPIIT UNAM IA104325, IN-106823, and IN-119123.      

Funding for the SDSS, SDSS-II, SDSS-V has been provided by the Alfred P. Sloan Foundation. The SDSS Web Site is \url{http://www.sdss.org/}.
SDSS acknowledges support and resources from the Center for High-Performance Computing at the University of Utah. The SDSS is managed by the Astrophysical Research Consortium for the Participating Institutions. 
The Participating Institutions are the National Science Foundation, the U.S. Department of Energy, the National Aeronautics and Space Administration, the Japanese Monbukagakusho, the Max Planck Society, the Higher Education Funding Council for England, the American Museum of Natural History, Astrophysical Institute Potsdam, University of Basel, University of Cambridge, Case Western Reserve University, University of Chicago, Drexel University, Fermilab, the Institute for Advanced Study, the Japan Participation Group, Johns Hopkins University, the Joint Institute for Nuclear Astrophysics, the Kavli Institute for Particle Astrophysics and Cosmology, the Korean Scientist Group, the Chinese Academy of Sciences (LAMOST), Los Alamos National Laboratory, the Max-Planck-Institute for Astronomy (MPIA), the Max-Planck-Institute for Astrophysics (MPA), New Mexico State University, Ohio State University, University of Pittsburgh, University of Portsmouth, Princeton University, the United States Naval Observatory, and the University of Washington.

\section*{Data Availability}

Optical spectra and spectral parameters used in this work as control samples are from the SDSS and can be downloaded from the Data Release 7 \citep{shenetal11}.

Optical spectra of the BZQ sample can be downloaded from the Data Release 16 \citep{ahumadaetal20}. The spectral parameters of the BZQ sample are provided as part of this paper in ascii format. For each source, {the redshift}, full width at half-maximum of \mgiionly, equivalent widths of \mgiionly\ and \feiiUV, and {the \mgiionly\ line luminosities} and their uncertainties are reported in units of km s$^{-1}$, \AA, and erg s$^{-1}$, respectively.



\bibliographystyle{mnras}
\bibliography{BZQmain} 




\appendix

\section{Statistical tests} \label{app:stat_tests}

This appendix provides the statistical tests employed in this work. 

When the assumption of normality is not satisfied for a distribution of values, non-parametric tests provide a robust alternative for comparing distributions. In this work, we consider the Kolmogorov--Smirnov (K-S) test \citep{KS51}, the Anderson--Darling (A-D) test \citep{AD54}, and the Mann--Whitney U (MWU) test \citep{Hollander14}.

The K-S test compares the empirical cumulative distribution functions (ECDFs) of two samples and is sensitive to differences in both location and shape. The A-D test extends this approach by applying a weighting that enhances sensitivity to differences in the tails of the distributions, which is particularly relevant for skewed or heavy-tailed data \citep[e.g.,][]{houetal09}.

For the two-sample A-D test, the statistic is defined as \citep{scholz87}
\begin{equation}
A^2 = \frac{nm}{N} \int_{-\infty}^{\infty}
\frac{\left[F_n(x) - G_m(x)\right]^2}
{H_N(x)\left[1 - H_N(x)\right]} \, dH_N(x),
\end{equation}
where $n$ and $m$ are the sizes of the two samples, $N = n + m$, $F_n(x)$ and $G_m(x)$ are their respective ECDFs, and $H_N(x)$ is the combined ECDF. This formulation allows a non-parametric assessment of whether the two samples are drawn from the same parent distribution, without assuming normality or equal variances \citep{stephens74}.

The MWU test provides a complementary non-parametric approach based on rank ordering rather than cumulative distributions. It evaluates whether values from one sample tend to be systematically larger or smaller than those from the other, making it particularly sensitive to differences in central tendency. Unlike the K-S and A-D tests, which compare entire distributions, the MWU test probes differences in the typical (median) values without assuming any specific functional form. This property makes it well suited for assessing shifts in line widths or emission-line strength parameters when distributions may overlap but differ in their typical values.

For all statistical tests, significance is assessed using the corresponding \textit{p}-values, which quantify the probability of obtaining a test statistic at least as extreme as the observed one under the null hypothesis that the two samples are drawn from the same parent distribution. Throughout this work, smaller \textit{p}-values indicate stronger evidence against the null hypothesis, with conventional significance thresholds adopted for reference as $0.05$.

\section{Selection of a representative sub-sample of radio non-detected AGN} \label{app:QSO_subsample}

This appendix describes the procedure adopted to construct a statistically representative sub-sample of the radio non-detected AGN population (QSO), following the methodology outlined in Appendix C of \citet{patinoetal26}.

The parent QSO sample comprises $N = 65{,}378$ sources, resulting in a strong imbalance with respect to the RL and blazar samples considered in this work. To mitigate potential biases arising from highly unequal sample sizes in two-sample statistical tests, we adopt a bootstrap-based sub-sampling strategy designed to preserve the statistical properties of the parent distribution while enabling controlled comparisons.

\subsection{Determination of the optimal sub-sample size}

We explored candidate sub-sample sizes in the range $n = 100$ to $n = 1000$. For each candidate size, a large number of random sub-samples were drawn \emph{without replacement}, ensuring that each realization consisted of unique objects and avoiding artificial duplication of data points. Each realization was evaluated using a set of complementary distributional similarity metrics that quantify how well the sub-sample reproduces the statistical properties of the full parent population. 

These metrics probe different aspects of the distribution, including global shape, central tendency, dispersion, and higher-order moments. To avoid reliance on any single metric, the results from all metrics and all realizations were combined (“stacked”), and the frequency with which each candidate size was identified as the most representative was recorded.



\section{Comparison of the \FWHMmg\ and \rfeUV\ Distributions}
\label{app:comp_metrics}

\begin{table*}
\centering
\caption{Statistical test results ($p$-values) for the \FWHMmg\ parameter obtained using the five representative QSO subsamples ($n\approx600$). QSO subsamples were selected using bootstrap resampling based on five similarity metrics: empirical cumulative distribution function (ECDF), Wasserstein distance, standardized mean difference, standard deviation ratio, and higher-moment differences.}
\label{tab:pvalues_FWHM}
\begin{tabular}{lccccc}
\toprule 
\textbf{Test} & \textbf{ECDF} & \textbf{Wasserstein} & \textbf{Std.\ Mean Diff.} & \textbf{StdDev Ratio} & \textbf{High Moment Diff.} \\
\midrule
\multicolumn{6}{l}{\textbf{BZQ full sample}}\\
K-S & $1.21\times10^{-4}$ & $3.30\times10^{-5}$ & $5.21\times10^{-5}$ & $2.31\times10^{-6}$ &$ 1.26\times10^{-8}$\\
A-D & 0.001 & 0.001 & 0.001 & 0.001 & 0.001\\
MWU & $5.21\times10^{-5}$ & $3.58\times10^{-5}$ & $2.26\times10^{-5}$ & $1.03\times10^{-5}$ & $7.78\times10^{-9}$\\
\midrule
\multicolumn{6}{l}{\textbf{BZQ-C}}\\
K-S & $2.02\times10^{-10}$ & $1.61\times10^{-10}$ & $8.99\times10^{-11}$ & $2.27\times10^{-11}$ & $2.50\times10^{-13}$\\
A-D & 0.001 & 0.001 & 0.001 & 0.001 & 0.001\\
MWU & $2.26\times10^{-10}$ & $1.65\times10^{-10}$ & $6.41\times10^{-11}$ & $6.73\times10^{-11}$ & $2.90\times10^{-13}$\\
\midrule
\multicolumn{6}{l}{\textbf{FRII-Q}}\\
K-S & $9.33\times10^{-21}$ & $3.39\times10^{-20}$ & $4.90\times10^{-21}$ & $2.75\times10^{-18}$ & $4.34\times10^{-15}$\\
A-D & 0.001 & 0.001 & 0.001 & 0.001 & 0.001\\
MWU & $9.30\times10^{-22}$ & $1.90\times10^{-21}$ & $1.75\times10^{-21}$ & $2.90\times10^{-20}$ & $1.40\times10^{-15}$\\
\midrule
\multicolumn{6}{l}{\textbf{BZQ-C vs FRII-Q (direct comparison)}}\\
K-S & $4.20\times10^{-30}$ & -- & -- & -- & --\\
A-D & 0.001 & -- & -- & -- & --\\
MWU & $1.10\times10^{-29}$ & -- & -- & -- & --\\
\bottomrule
\end{tabular}
\\
\textbf{Notes.}
The A-D probabilities reported as $p=0.001$ correspond to the lower numerical limit returned by the \texttt{scipy.stats.anderson\_ksamp} implementation. The direct BZQ-C versus FRII-Q comparison is independent of the QSO subsample selection criterion and is therefore reported only once.
\end{table*}

\begin{table*}
\centering
\caption{Statistical test results ($p$-values) for the \rfeUV\ parameter obtained using the five representative QSO subsamples ($n\approx600$).}
\label{tab:pvalues_RFeII}
\begin{tabular}{lccccc}
\toprule
\textbf{Test} & \textbf{ECDF} & \textbf{Wasserstein} & \textbf{Std.\ Mean Diff.} & \textbf{StdDev Ratio} & \textbf{High Moment Diff.}\\
\midrule
\multicolumn{6}{l}{\textbf{BZQ full sample}}\\
K-S & $1.75\times10^{-6}$ & $2.79\times10^{-6}$ & $2.48\times10^{-5}$ & $1.48\times10^{-5}$ & $3.58\times10^{-7}$\\
A-D & 0.001 & 0.001 & 0.001 & 0.001 & 0.001\\
MWU & $3.73\times10^{-6}$ & $1.66\times10^{-6}$ & $2.81\times10^{-5}$ & $3.57\times10^{-5}$ & $2.20\times10^{-8}$\\
\midrule
\multicolumn{6}{l}{\textbf{BZQ-C}}\\
K-S & $3.87\times10^{-2}$ & $3.82\times10^{-2}$ & $5.74\times10^{-2}$ & $1.07\times10^{-1}$ & $1.80\times10^{-2}$\\
A-D & $1.86\times10^{-2}$ & $1.48\times10^{-2}$ & $3.13\times10^{-2}$ & $4.31\times10^{-2}$ & $4.05\times10^{-3}$\\
MWU & $3.40\times10^{-2}$ & $2.59\times10^{-2}$ & $6.30\times10^{-2}$ & $7.65\times10^{-2}$ & $5.91\times10^{-3}$\\
\midrule
\multicolumn{6}{l}{\textbf{FRII-Q}}\\
K-S & $1.66\times10^{-7}$ & $4.52\times10^{-7}$ & $2.02\times10^{-6}$ & $1.25\times10^{-6}$ & $3.35\times10^{-8}$\\
A-D & 0.001 & 0.001 & 0.001 & 0.001 & 0.001\\
MWU & $5.11\times10^{-7}$ & $2.57\times10^{-7}$ & $1.99\times10^{-6}$ & $4.77\times10^{-6}$ & $8.08\times10^{-9}$\\
\midrule
\multicolumn{6}{l}{\textbf{BZQ-C vs FRII-Q (direct comparison)}}\\
K-S & 0.0843 & -- & -- & -- & --\\
A-D & 0.0429 & -- & -- & -- & --\\
MWU & 0.1595 & -- & -- & -- & --\\
\bottomrule
\end{tabular}
\end{table*}

This Appendix demonstrates that the statistical conclusions presented in the main text are insensitive to the particular representative QSO subsample adopted. Independent sets of representative QSO subsamples {($n\approx600$)} were constructed for the \FWHMmg\ and \rfeUV\ analyses using the five bootstrap similarity metrics described in Section~\ref{ssec:shen_subsample}: (i) maximum ECDF difference, (ii) Wasserstein distance, (iii) standardized mean difference, (iv) standard deviation ratio, and (v) higher-order moment differences.

For each parameter, the bootstrap realization providing the closest match to the parent QSO distribution under each metric was selected as the representative subsample. The resulting QSO subsamples were then compared with the radio-loud populations using the K-S, A-D, and MWU tests. Tables~\ref{tab:pvalues_FWHM} and \ref{tab:pvalues_RFeII} summarize the corresponding $p$-values.

Overall, the three non-parametric tests provide mutually consistent results across all five representative QSO subsamples. For \FWHMmg, statistically significant differences are recovered for every QSO selection criterion and for the direct BZQ-C--FRII-Q comparison, demonstrating that the inferred population ordering is insensitive to the definition of the reference QSO sample. In contrast, the \rfeUV\ analysis yields systematically weaker statistical evidence, and the direct comparison between BZQ-C and FRII-Q is not significant in the K-S and MWU tests, although the A-D test indicates a marginal difference. These results confirm that the principal population separation in the UV Eigenvector~1 plane is driven by \FWHMmg, while \rfeUV\ provides a weaker secondary discriminator.


\bsp	
\label{lastpage}
\end{document}